\documentclass[12pt]{article}
\usepackage{amsfonts,amsmath}
 \usepackage{amssymb}
 \usepackage[hypertex] {hyperref}

\makeatletter \@addtoreset{equation}{section} \makeatother

\newcommand{\be}{\begin{equation}}
\newcommand{\ee}{\end{equation}}
\newcommand{\bee}{\begin{eqnarray}}
\newcommand{\beee}{\begin{array}}
\newcommand{\eee}{\end{eqnarray}}
\newcommand{\eeee}{\end{array}}

\newcommand{\un}{{\underline{n}}}

\newcommand{\ga}{\alpha}
\newcommand{\pa}{{\dot{\ga}}}

\newcommand{\gb}{\beta}
\newcommand{\gga}{\gamma}

\newcommand{\W}{{\cal W}}
\newcommand{\K}{{\cal K}}
\newcommand{\F}{{\cal F}}
\newcommand{\Ll}{{\cal L}}

\newcommand{\Gg}{{\cal G}}
\newcommand{\T}{{\cal T} }
\renewcommand{\S}{{\cal S}}
\newcommand{\Q}{{\cal Q} }
\newcommand{\rhs}{{\it r.h.s.} }

\newcommand{\ie}{{\it i.e.,} }
\newcommand{\ls}{\!\!\!\!\!\!}
\def\ck{{ K}}

\newcommand{\gvep}{\varepsilon}

\newcommand{\go}{\omega}

\newcommand{\by}{{\bar{y}}}

\newcommand{\q}{\,,\qquad}

\newcommand{\dga}{{\dot{\alpha}}}
\newcommand{\dgb}{{\dot{\beta}}}

\newcommand{\nn}{\nonumber}

\newcommand{\half}{\frac{1}{2}}

\newcommand{\p}{\partial}

\newcommand{\f}{\frac}
\newcommand{\B}{{\cal B}}

\newcommand{\R}{{\cal R}}

\newcommand{\U}{\Upsilon}
\newcommand{\ups}{\upsilon}
\newcommand{\bu}{\bar{\upsilon}}

\def\S{\Sigma}

\def\ck{{\Upsilon}_V}
\def\cl{{\cal L}}

\newcommand{\dr}{{\rm d}}

\newcommand{\xx}{{\bf x}}
\newcommand{\zz}{{\bf z}}

\newcommand{\Sp}{{\mathcal H}}
\newcommand{\Spl}{{\mathcal H}^{loc}}
\newcommand{\pp}{\partial^*_Z{}}

\begin{document}

\begin{flushright}
{\small FIAN/TD/02-15}
\end{flushright}
\vspace{1.7 cm}

\begin{center}
{\large\bf Invariant Functionals in  Higher-Spin Theory }

\vspace{1 cm}

{\bf  M.A.~Vasiliev}\\
\vspace{0.5 cm}
{\it
 I.E. Tamm Department of Theoretical Physics, Lebedev Physical Institute,\\
Leninsky prospect 53, 119991, Moscow, Russia}

\end{center}

\vspace{0.4 cm}

\begin{abstract}
\noindent
A new construction for  gauge invariant functionals in the nonlinear higher-spin theory is proposed.  Being supported by differential forms closed  by virtue of
the higher-spin equations, invariant functionals  are associated with  central
elements of the higher-spin algebra. In  the on-shell $AdS_4$ higher-spin theory
we identify a four-form conjectured to represent the generating functional for
$3d$ boundary correlators and a two-form  argued to support charges for black hole
solutions. Two actions for  $3d$ boundary conformal higher-spin theory are
associated with the two parity-invariant higher-spin models in $AdS_4$.
The peculiarity of the spinorial formulation of the on-shell $AdS_3$ higher-spin
theory, where the invariant functional is supported by a two-form,
is conjectured to be related to the holomorphic factorization at the boundary.
The nonlinear part of the star-product function $F_*(B(x))$ in the  higher-spin
equations is argued to lead to divergencies in the  boundary limit representing
  singularities at coinciding boundary space-time points of the factors of
$B(x)$, which can be regularized by the point splitting.
An interpretation of the RG flow in terms of proposed construction is briefly discussed.

\end{abstract}

\newpage
\tableofcontents

\newpage

\section{Introduction}
\label{intro}

Standard holographic prescription for computation of correlators
\cite{Maldacena:1997re,Gubser:1998bc,Witten:1998qj,D'Hoker:1998tz} is based on the
bulk action evaluated on  solutions of the dynamical field
equations with appropriate boundary conditions. The lower-order
actions for higher-spin (HS) gauge fields, that extend the Fronsdal's
quadratic actions \cite{Frhs,Frfhs} to the cubic order, are known
since \cite{Bengtsson:1983pd,Berends:1984wp,Fradkin:1987ks} (see
also \cite{Vasilev:2011xf,Joung:2011ww,Bekaert:2014cea} for recent
progress and more references).
These actions are however incomplete even at the cubic order, not
fixing relative coupling constants of  cubic vertices. The
lower-order results indicate that the full nonlinear extension of
the Fronsdal's action does exist. However, unavailability of its
explicit form complicates the holographic analysis of the HS
theories. An interesting alternative proposal suggested in
\cite{Boulanger:2011dd,Sezgin:2011hq}, where the action is
defined in a higher-dimensional space-time, leads,
however, to unconventional actions even for lower spins and its
application in the context of HS holography remains to be
explored.

Despite the impressive progress on the verification of the Klebanov-Polyakov conjecture
\cite{Klebanov:2002ja,Leigh:2003gk,Sezgin:2003pt}
on the holographic duality between HS gauge theories and vectorial boundary
theories achieved via analysis of the HS field equations in \cite{Giombi:2009wh}
(for more references and recent developments see, e.g.,
\cite{Vasiliev:2012vf,Maldacena:2012sf,Giombi:2012ms,Colombo:2012jx,Didenko:2012tv,
Jevicki:2012fh,Giombi:2013fka,
Giombi:2014yra,Beccaria:2014jxa,Koch:2014aqa,Giombi:2014xxa,Beccaria:2014xda,
Barvinsky:2014kta}),
it is desirable to have a direct prescription for the generating functional
of boundary correlators. The situation with the $AdS_3/CFT_2$ HS holography
\cite{Henneaux:2010xg,Campoleoni:2010zq,Gaberdiel:2010pz} is analogous.

In this paper we propose a new approach to the construction of
invariant functionals in HS theory which leads to  differential forms $L(\phi)$ built from dynamical fields $\phi$, that are closed,
\be
\dr L(\phi)=0\,,
\ee
by virtue of the  nonlinear HS field equations. To this end we suggest the
extension of the nonlinear HS field equations of \cite{more} which determines
invariant densities $L(\phi)$ associated with central elements of the HS algebra.
The functional
\be
\label{Sact}
S=\int L(\phi)
\ee
turns out to be gauge invariant.

Generally, there exist two types of unfolded systems called {\it off-shell}
 and {\it on-shell}. Off-shell systems
describe a set of constraints that express a (usually infinite)
set of auxiliary fields via derivatives of some ground fields imposing no
differential restrictions on the latter. On-shell systems impose
differential field equations on the ground fields called dynamical
in this case. For off-shell HS systems  the
 functional $S$ is anticipated to describe the action.
For on-shell systems $S$ can be thought of as an on-shell action underlying the
analysis of $AdS/CFT$. In this paper we focus on the on-shell spinorial
HS theories in $AdS_4$ and $AdS_3$.

In the standard $AdS/CFT$, the generating functional of the boundary theory on
$\Sigma$ is identified with \cite{Gubser:1998bc}
\be
\label{Se}
S_\epsilon = \int_\epsilon^\infty  d\zz \int_{\Sigma}L(\phi)\,
\ee
as a functional of appropriate boundary values of fields.
Here $\zz$ is the Poincar\'e coordinate integrated till the cutoff $\epsilon$.
In this setup,  the cutoff can break the symmetries at the boundary
while the generating functional can depend on total derivatives in $L$.
In principle, the latter can be adjusted to ensure appropriate properties of the theory
in spirit of, e.g., \cite{Arutyunov:1998ve,Metsaev:2009ym,Metsaev:2014vda}.

Alternatively,  we suggest to consider the  functional of the form
\be
\label{S}
S=\f{1}{2\pi i}\oint_{\zz=0} \int_{\Sigma}L(\phi)\,
\ee
resulting from the integration over a cycle on the plane of complexified
$\zz$  encircling the infinite point $\zz=0$.
As explained in \cite{Vasiliev:2012vf} (see also below),
the possibility of the integration in the plane of complex $\zz$ is
provided by the unfolded formulation of HS field equations operating with differential forms and allowing at least locally to extend   the system
to a larger space including the complexified one. In this construction,
$L(\phi)$ remains closed in the extended space and $S$ remains invariant
under all gauge symmetries of the original bulk system provided that the pullback
 is well defined in some neighborhood of the infinity $\zz=0$
allowing the integration around  $\zz=0$. In the absence of branch cuts
 the resulting functional is real coinciding with the residue with respect to $\zz$ of the
 original real function of real $\zz$.

Note  that if the system was off-shell in the original
space, its dynamical fields will necessarily obey certain differential equations
 in the extended space allowing $L$ be closed in a larger space.\footnote{This is somewhat analogous to  the Group Manifold Approach
\cite{Castellani:1981um,Castellani:1991ev} (see also \cite{Hu:2015cwa})
requiring so-called rheonomy conditions on the curvatures  to extend  the system to
the higher-dimensional  group manifold.} If the extension to
the complex plane of $\zz$ exhibits branch cuts  the standard definition (\ref{Se})
may be more appropriate. This is unlikely  to happen
in the $AdS_4$ HS model but remains to be investigated in other models.

Though the whole setting also applies to the standard construction  (\ref{Se}),
for definiteness, in the sequel we will mostly refer to  boundary functionals
(\ref{S}).  As argued in Section \ref{Parity}, in certain cases functionals
(\ref{S}) describe local actions for boundary conformal HS theory that only
give local contribution to the boundary correlators. To reproduce the nonlocal
part of the correlators, in these cases one should either use the standard
construction  (\ref{Se}) or a limiting procedure explained in Section
 \ref{Parity}.

Let us consider the $AdS_4/CFT_3$ case in some more detail.
In  spinor notation with two-component spinor indices $\ga,\gb=1,2$,
$\dga,\dgb =1,2$, local coordinates of $AdS_4$  are
\be
\label{xz}
x^{\ga\dga }=(\xx^{\ga\dga},-\f{i}{2} \epsilon^{\ga\dga} \zz^{-1})\,,
\ee
where the symmetric part of $4d$  coordinates
$\xx^{\ga\dga}=\xx^{\dga\ga}$ is identified with coordinates of
the boundary $\Sigma$ while
$\zz^{-1}$ is the radial coordinate of $AdS_4$. The appearance of
$\epsilon^{\ga\dga}=-\epsilon^{\dga\ga}$ in the definition of $\zz$ breaks the $4d$ Lorentz
symmetry $sp(2;\mathbb{C})$ to the $3d$ Lorentz symmetry
$sp(2;\mathbb{R})$ which acts on the both types of spinor
indices. In Poincar\'e coordinates, $AdS_4$ vierbein and Lorentz connection
can be chosen in the form
\be\label{poframe}
e^{\ga\dga} = \f{1}{2\zz} dx^{\ga\dga}\q \go^{\ga\gb}=-
\f{i}{4\zz} d\xx^{\ga\gb}\q \bar \go^{\dga\dgb} =
\f{i}{4\zz} d\xx^{\dga\dgb}\,.
\ee
Meromorphic dependence on $\zz$ makes it possible to complexify the
Poincar\'e coordinate $\zz$. The connection (\ref{poframe}) remains flat provided that all its $ d\bar \zz$ components  are  zero.
Strictly speaking, this is true everywhere except for the point $\zz=0$ of infinity since
$\f{\p}{\p \bar \zz} \f{1}{\zz}\neq 0$. Hence, our analysis  applies to the
complexified (Poincar\'e patch of) $AdS_4$ space with removed infinity $\zz=0$.

 In the $AdS/CFT$ correspondence dictionary, the
source term for the spin-$s$ conserved current $J_{n_1\ldots n_s}$ is
\be
\label{cop}
S = \int dx^3 \varphi^{n_1\ldots n_s} J_{n_1\ldots n_s}\,.
\ee
The current conservation
\be
\p_m J{}^m{}_{n_2\ldots n_{s}}=0
\ee
 is dual to the gauge symmetry of the gauge field
 \be\delta \varphi_{n_1\ldots n_s}=
\p_{(n_1}\gvep_{n_2\ldots n_{s})}\,.
\ee
The variation of $\langle\exp{-S}\rangle$ over $\varphi_{n_1\ldots n_s}$
gives  correlators of currents.

In the frame-like approach to $3d$ boundary theory, the symmetric tensor
field $\varphi_{n_1\ldots n_s}$ is substituted by the frame-like one-form
connection
$\go_{\ga_1\ldots \ga_{2(s-1)}}=dx^{\bf \un} \go_{{\bf \un}\,\ga_1\ldots \ga_{2(s-1)}}$
where the indices $\ga=1,2$ are spinorial and ${\bf \un}=0,1,2$. As explained in
\cite{Vasiliev:2012vf},
the role of $J_{n_1\ldots n_s}$ is played by the so-called HS Weyl tensor
$C_{\ga_1\ldots \ga_{2s}}$ and its conjugate $\bar C_{\dga_1\ldots \dga_{2s}}$. Though  $C_{\ga_1\ldots \ga_{2s}}$ is not
an operator at the boundary but rather the gauge-invariant curvature tensor
built in terms of $s$ derivatives of the connections $\go_{\ga_1\ldots \ga_{2(s-1)}}$,
it obeys the conservation condition (which from the bulk  perspective is
the Bianchi identity) and is a primary field of the conformal module
equivalent to that of the $3d$ conformal current. The counterpart of
action (\ref{cop}) is
\be
\label{freel}
S_2(\go)=\int L^4_2 \q L^4_2=\half
\go_\xx{}^{\ga_1\ldots \ga_{2(s-1)}} e_{\zz}
e_\xx{}^{\ga_{2s-1}}{}_\gga e_\xx^{\ga_{2s} \gga}
( a C_{\ga_1\ldots \ga_{2s}}(\go)+\bar a \bar C_{\ga_1\ldots \ga_{2s}}(\go))\,,
\ee
where $e_{\xx\,\ga\gb}$ is the boundary dreibein one-form,
 $e_{\zz }$ is the component of the $AdS_4$  vierbein along the Poincar\'e coordinate
and $a$, $\bar a $ are some complex conjugate coefficients
(in the sequel the wedge symbol is omitted since all products are wedge products).

The functional $S_2(\omega)$  represents the $\omega$-dependent quadratic part
of the full invariant functional (\ref{S}) where the coefficients $a$ and $\bar a $
should be determined by the explicit computation in a particular HS model.
Depending on a model,
different linear combinations of the Weyl zero-forms represent either $R_{\xx\xx}$ or
$R_{\xx\zz}$ components of the HS curvatures at $\zz=0$
\be
\label{Rxxzz}
R_{\xx\xx}\sim a^{loc} e_\xx e_\xx C+
\bar a^{loc} e_\xx e_\xx \bar C\q
R_{\xx\zz}\sim a^{nloc} e_\zz e_\xx  C+
\bar a^{nloc} e_\zz e_\xx \bar C\,.
\ee
(For explicit expressions see \cite{Vasiliev:2012vf}.)

At the quadratic level, the decomposition of  invariant functional (\ref{freel})
into the local and nonlocal parts
\be
\label{locnon}
S=S^{loc} +S^{nloc}
\ee
corresponds to the decomposition of (\ref{freel}) into a linear combination of
the terms proportional to $R_{\xx\xx}$ and $R_{\xx\zz}$, respectively.
Indeed, the part of $S$ that only contains  the boundary derivatives of
boundary fields describes some boundary functional.
Correspondingly,  (\ref{freel}) with  $a=a^{loc}$, $\bar a =\bar a^{loc}$
describes the boundary Chern-Simons
action of conformal HS theory {\it a la}
\cite{Pope:1989vj,Fradkin:1989xt,Horne:1988jf,Nilsson:2013tva}. This
gives  local contribution to the correlators. In the  case $a=a^{nloc}$, $\bar a =\bar a^{nloc}$,
the action (\ref{freel}) contains the bulk derivative hence
giving a generating function for the nonlocal part of the  correlators.
As explained  in Section \ref{Parity}, for the $P$-invariant HS models
 the  naive functional $S$ (\ref{S}) gives rise to the local
 boundary conformal HS theory with $S^{nloc}=0$ while $S^{nloc}$ can be associated
 with its derivative over the parameter $\eta$ in the nonlinear HS equations.
It should be stressed that with this definition the resulting local functional on
the boundary is $P$-odd while the nonlocal one is $P$-even.

In addition to the HS conformal gauge fields, the model contains
spin-zero conformal currents  of different conformal dimensions
\be
C_1(\xx,\zz)=C(\xx,\zz)\q C_2 (\xx,\zz)=\p_\zz C(\xx,\zz)\,.
\ee
In this sector, the action functional is
\be
\label{freel0}
S_2(C)=\int L_2 \q L_2= V_\Sigma e_{\zz}  C_1(x) C_2(x)\,,
\ee
where $V_\Sigma$ is the $3d$ volume form.
The behavior of the fields $C(\xx,\zz)$ and $\go(\xx,\zz)$ at $\zz\to 0$
is in agreement with their conformal dimensions (for more detail see \cite{Vasiliev:2012vf}).
As a result,  the generating functional (\ref{freel0}) just singles out
the conformal invariant part of  $L_2$.

Let the dynamical boundary fields, which are the primary conformal components
among $\go$ and $C_1$ or $C_2$, be collectively denoted  $\phi(x)$. Then the
boundary correlators are conjectured to be given by
\be\label{corr}
\langle J(\xx_1)J(\xx_2)\ldots \rangle = \f{\delta^n \exp{[-S
(\phi)]}}{\delta \phi(x_1)
\delta\phi(x_2)\ldots} \Big |_{\phi=0}\,,
\ee
where
\be
x_1 = (\xx_1\,, \zz)\q x_2 = (\xx_2\,, \zz)\q\ldots
\ee
are taken at different boundary points $\xx_1$, $\xx_2$, $\ldots$ and some
small $\zz$ inside the integration contour in the definition (\ref{S}).
(The issue of the dependence on $\zz$ is  analogous to that
\cite{Gubser:1998bc} in the standard approach (\ref{Sact}).)

Evaluation of (\ref{corr}) for $\phi=C_1$ or $\phi=C_2$ is equivalent to the
evaluation of correlators with different boundary conditions in the standard
approach, namely with $C_2=0$ or $C_1=0$, respectively.
Analogously, one can choose the generalized
Weyl tensor as an independent field, expressing connections $\go$ in terms of $C$
by the field equations (though the resulting expressions are nonlocal and are defined
modulo the gauge freedom). Combining
two such exchanges with the dualization of the HS Weyl tensors should reproduce
 the Witten's $SL(2,\mathbb{Z})$ duality \cite{Witten:2003ya} extended to
higher spins by Leigh and Petkou \cite{Leigh:2003ez}.

The problem is to find  a density  $L$ leading to the
gauge invariant functional $S$ (\ref{S}) in the full nonlinear HS theory.
In this paper we mostly focus on the general scheme
which opens a new way toward  solution of  this problem. Main attention will be payed
to the spinorial $AdS_4$ HS theory where in particular we identify the local boundary
functionals which are anticipated to describe $3d$ conformal HS theories and are
associated with the so-called $A$ and $B$ HS theories. Also we briefly consider the on-shell spinorial HS
model in $2+1$ dimensions. Elaboration of the detailed structure of the invariant
functionals introduced in this paper  requires significant technical
work to be presented elsewhere \cite{Didenko:2015cwv,DMV}.

Apart from invariants associated with the density  forms  of maximal degree,
our construction gives rise  the
on-shell densities of lower degrees. In particular, the $3d$ and $4d$ on-shell HS systems considered in
this paper admit the closed two-form $L^2$. For the $4d$ HS system this is conjectured
to describe the black hole (BH) charge  as
\be
\label{entr}
Q\sim \int_{\Sigma^2} L^2
\ee
integrated over a cycle $\Sigma^2$ surrounding a BH singularity. Since $L^2$ is closed,
$Q$ is insensitive to local variations of $\Sigma^2$. Hence,
 $\delta Q$ evaluated at infinity equals to  $\delta Q $ evaluated at the BH horizon.
Assuming that thermodynamical first law is to hold true for HS BHs,  it must be
controlled by this relation.  To make contact with the
standard approach \cite{Wald:1993nt} (for more detail see \cite{Didenko:2015pjo})
one should take into account some novelties of our construction.

First of all, it may look surprising that
the two-form $L^2$ exists at all since it is closed and gauge invariant up to exact
forms not just for a BH solution that admits
Killing vectors but for any solution including, in particular, fluctuations around the
BH solution. Here it is important that $L^2$ is not a  local functional
of fields. Rather it is (minimally) nonlocal in the sense specified in  \cite{Vasiliev:2015wma},
depending on all derivatives of the fields
and containing inverse powers of the background curvature in the derivative expansion
that, in particular, complicates a straightforward flat limit analogously to the situation
with the HS actions \cite{Fradkin:1987ks}. Nontheless
 $L^2$ should be well defined as a space-time  closed form,
\ie  (\ref{entr}) makes sense for any $\Sigma^2$.

The infinity cycle $\Sigma_\infty^2$ and the horizon cycle $\Sigma_{H}^2$ are special. At $\Sigma_\infty^2$, where the theory becomes asymptotically
free and $L^2$ becomes asymptotically local, $Q$  reproduces usual asymptotic charges
\cite{Didenko:2015pjo}. (It would be interesting to establish their explicit relation to
the construction of \cite{Barnich:2001jy,Barnich:2005bn}.)
The horizon $\Sigma_{H}^2$ is  a Killing bifurcation surface.
As discussed in Section \ref{Black holes} for the case of GR,
from the perspective of unfolded equations
this implies trivialization of the evolution equations in certain directions. So far
it is not known whether or not a horizon $\Sigma_{H}^2$ possessing such properties
can be associated with the HS solutions of
\cite{Didenko:2009td,Iazeolla:2011cb,Bourdier:2014lya} to the full nonlinear
HS equations. To answer this question it should be explored whether there exists such a surface
$\Sigma_{H}^2$ on which some of the unfolded equations trivialize in terms of
the coordinates of the observer at infinity. For $L^2 \big |_H$, starting with the volume
form on $H$ times a  constant proportional to $\beta$, $Q$
will start with the term proportional to the area of $H$.

Hopefully, the realization of the BH charge in terms of $L^2$ $(L^{d-2}$ for higher
dimensions) can help to clarify the microscopic origin of the BH entropy
the profound example of which was proposed in \cite{Strominger:1996sh}.
A natural guess is to identify the Lagrangian
of the microscopic system with $L^2 \big |_H$ for the restriction of
the original unfolded system  to the  horizon $H$.

As discussed in Section \ref{ads3}, the $3d$ on-shell HS system, where the only
invariant density  is a two-form $L^2$, is special.  Naively the form degree two is
smaller than anticipated for a boundary generating function  and larger than is needed for the
$3d$ BH charge. However, very likely this is  just appropriate for the
both problems with the invariant functionals of the form
\be
S = \int_{S^1\times \Sigma^1} L^2\,,
\ee
where $S^1$ is a cycle around $AdS_3$ infinity as in (\ref{S}) while $\Sigma^1$
is either a cycle at the conformal boundary for the generating functional of
boundary correlators or a cycle around the singularity of the BTZ-like BH solutions
\cite{BTZ}, which in the HS theory were considered in \cite{Didenko:2006zd,Ammon:2012wc}
(and references therein).

Since it is hard to consider in detail all these
questions in a single paper, here we focus on the
general scheme providing a starting point for the future studies.
The rest of the paper is organized as follows. In Section \ref{fda}
we summarize general properties of invariant functionals in the
unfolded dynamics approach and related interpretation of the RG
flow. The structure of the field equations of the nonlinear HS theory in $AdS_4$
is recalled  in Section \ref{Nonlinear Higher-Spin Equations}.
Subtleties of the boundary limit in HS theories affecting conformal properties
 in their holographic interpretation are also discussed here.
In particular  it is shown that any nonlinear
star-product function of the zero-form $B(x)$ in the nonlinear HS
equations exhibits divergencies in  the boundary limit. It is argued however that
these divergencies represent
  singularities at coinciding boundary space-time points of the factors of
$B(x)$, that can  be regularized by  the point splitting.
General structure of the
extended unfolded systems allowing to define invariant densities
and its application to the $AdS_4$ HS theory are presented in
Sections \ref{struc} and \ref{invfunct}, respectively. In particular,
possible application to BH physics is  sketched in Section \ref{Black holes}.
The on-shell $AdS_3$ spinorial HS theory is considered in
Section \ref{ads3}.
 Section \ref{Conclusion} contains conclusions.

\section{Unfolded equations and invariant functionals}
\label{fda}

Let $M^d$ be a $d$-dimensional  manifold (space-time)
with local coordinates $x^\un$ ($\un = 0,1,\ldots d-1$).
By unfolded formulation of a linear or nonlinear
system of partial differential equations
in $M^d$ we mean its reformulation in the first-order form
\cite{Ann}
\be
\label{unf} \dr_x W^\Omega (x)= G^\Omega (W(x))\,,
\ee
where $ \dr_x=dx^\un  \frac{\p}{\p x^\un}\, $
is the exterior  derivative in $M^d$, $W^\Omega(x)$
is a set of degree-$p_\Omega$ differential forms,
and $G^\Omega (W)$ is some degree-$(p_\Omega +1)$
function of $W^\Lambda$
\be
G^\Omega (W) =
\sum_{n=1}^\infty f^\Omega{}_{\Lambda_1\ldots \Lambda_n} W^{\Lambda_1} \ldots
 W^{\Lambda_n}\,
\ee
that satisfies the  generalized Jacobi identity on the structure coefficients
$f^\Omega{}_{\Lambda_1\ldots \Lambda_n}$
\be \label{BI} G^\Lambda (W) \f{\p
G^\Omega (W)} {\p W^\Lambda}  =0\,.
\ee
Strictly speaking,  generalized Jacobi
identities (\ref{BI}) have to be satisfied  at $p_{\Omega} < d$
since any $(d+1)$-form  in ${ M}^d$ is zero. Any solution of
(\ref{BI})  defines a free differential algebra
\cite{Sullivan,FDA,FDA1,FDA2}. A free differential algebra is  {\it
universal} \cite{Bekaert:2005vh,act} if (\ref{BI}) holds
 independently of the space-time dimension, \ie for abstract
 supercoordinates $W^\Lambda$ which are (anti)commuting
 for variables associated with differential forms of (odd)even degrees. All free
differential algebras associated with known HS theories  are universal.

Condition (\ref{BI}),  which can equivalently be written as
\be
\label{qdif}
Q^2 =0\,,\qquad Q:=  G^\Omega (W)  \f{\p}{\p W^\Omega}\,,
\ee
guarantees formal consistency of  unfolded system (\ref{unf})
which can be put into the Hamiltonian-like form
\be
\label{unf1}
\dr_x F(W(x)) =  Q (F(W(x))
\ee
for any $F(W)$. Universal equation (\ref{unf}) is invariant under the gauge transformation
\be \label{delw} \delta W^\Omega = \dr_x \varepsilon^\Omega +\varepsilon^\Lambda
\frac{\p G^\Omega (W) }{\p W^\Lambda}\,,
\ee
where the gauge parameter
$\varepsilon^\Omega (x) $ is a $(p_\Omega -1)$-form.  (Zero-forms
 have no  gauge parameters.)

Dynamics of a universal unfolded system
is characterized entirely by  differential $Q$ (\ref{qdif})
defined on the ``target space" of dynamical variables $W^\Omega$
independently of the original space-time. In particular,
invariants like actions and conserved
charges are characterized by  the $Q$--cohomology.
Indeed, as shown in \cite{act}, a gauge invariant functional
is an integral over a $p$-cycle $M^p$
\be \label{action1} S=\int_{M^p} \cl (W)\,
\ee
of some $Q$-closed  {\it Lagrangian $p$--form} $\cl(W)$
\be
\label{qcl}
Q\cl=0\,:\qquad G^\Omega (W)  \f{\p \cl(W)}{\p W^\Omega}  =0\,.
\ee
(It is elementary to see that such $S$ is
 invariant under gauge transformations (\ref{delw}).)
If $\cl$ is $Q$-exact, by virtue of (\ref{unf1})
it is $\dr_x$--exact giving a trivial functional up to possible
boundary terms. Hence nontrivial invariant functionals represent
$Q$-cohomology of the system in question.
Analysis of invariant functionals  in terms of $Q$-cohomology, which
applies to both linear and nonlinear unfolded systems (for examples see
\cite{act}), is complete:
any invariant functional of the universal unfolded field equations corresponds to some
their $Q$-cohomology. However, as for any other general approach, direct search of
invariant functionals via $Q$-cohomology may be  involved  for concrete
nonlinear systems.

The remarkable feature of  universal unfolded equations (\ref{unf}), which has deep
connection \cite{Vasiliev:2012vf} with holographic duality, is that they can be written
in space-times of different dimensions since the fact of their consistency is
insensitive to the number of space-time coordinates. Whether  unfolded system
(\ref{unf}) is on-shell or off-shell depends in the first place on the dimension
of space-time where it is considered. A typical situation
 is when the same unfolded system is off-shell in $d$ dimensions and
on-shell in $d+1$ dimensions.
If a system in $d$ dimensions is off-shell it can only have nontrivial
$H^d(Q)$-cohomology since in the topologically trivial case it is impossible
to construct a closed local functional of (derivatives) of the ground fields
not subjected to any field equations, that is not exact.\footnote{Note that this analysis
is local, discarding possible topological obstructions.
In particular, in this setup, topological invariants like Chern classes are
treated as locally exact.} Hence, actions $S^d$
of off-shell systems in $d$ dimensions are usually $d$-forms. However
$S^d$ remains $Q$-closed for the same dynamical system uplifted to
higher dimensions where it becomes on-shell.

The property that $\Ll$ is $Q$-closed suggests that  formula (\ref{S}) should
have general applicability  beyond HS gauge theories. Indeed, this
implies that $\Ll$ remains $\dr$-closed in a larger space with  complexified $\zz$.
As a result, $S$ (\ref{S})
turns out to be independent of local variations of the integration contour.

Note that the distinguished r\'ole of the closed
functionals in the unfolded dynamics
may also be related to the fact that averaging over integration cycles
if occurs in some underlying fundamental  theory
has no effect on the integrals of closed forms, giving zero for other
functionals. Since in the unfolded dynamics the gauge symmetries are consequences
of the $Q\sim \dr_x$ closure of the Lagrangian forms, it is tempting to
speculate that, other way around,  gauge symmetries can result from some sort
of averaging over integration cycles.\footnote{It should be noted that
even non-gauge systems like scalar field acquire gauge symmetries in their
unfolded form. These are gauge symmetries of the background one-form flat
connections  expressing coordinate independence of the unfolded formulation. }

Assuming that the infinity  is the only singular point and
choosing the contour around  $\zz=0$ to be a circle of radius $r$ on the
complex  $\zz$-plane, this implies
\be
\f{d S}{d r}=0\,.
\ee
By virtue of (\ref{unf1}) this is equivalent to
\be
\label{GS}
G_r^\Omega (W) \f{\p S(W)}{\p W^\Omega}=0\,,
\ee
where $G_r^\Omega$ is the component of $G^\Omega$ along the radial direction, \ie
discarding the terms not containing $dr$,
\be
G^\Omega= dr G^\Omega_r +\ldots\,.
\ee

In the perturbative analysis, the forms $W^\Omega$ are decomposed into
the vacuum part $W_0^\omega$ (the index $\omega$ is different from $\Omega$ to stress
that some vacuum components of $W^\Omega$ may be zero)
and the fluctuational part $W_1^\Omega$. Decomposing $G^\Omega (W)$ into the vacuum and fluctuational
parts
\be
G^\Omega (W) = G_0^\omega (W_0) + G^{\prime\Omega} (W_0, W_1)
\ee
at the condition that
\be
G^{\prime\Omega} (W_0, 0) =0\,,
\ee
guaranteeing that the vacuum fields do not source the dynamical ones,
 equation (\ref{GS}) can be  rewritten in the Hamiltonian-like form
\be
\label{ham}
\dot S(W_0,W_1) + H S(W_0,W_1)=0\,,
\ee
where
\be
\dot F(W_0,W_1) := G_{r 0}^\omega (W_0) \f{\p F(W_0,W_1)}{\p W_0^\omega}\,,
\ee
\be
H:=G_r^{\prime\Omega} (W_0, W_1) \f{\p}{\p W_1^\Omega}\,.
\ee
Note that the property that $W^\omega_0$ is a solution to (\ref{unf}) at
$W^{\prime \Omega}=0$ implies that
$G^{\prime\Omega}$ only contributes to the evolution of the fluctuations
$W^{\prime \Omega}$.

Using (\ref{unf}) for the vacuum solution
 it follows that $\dot F(W_0,W_1)$ indeed describes the $r$-evolution due to the
dependence of $W_0^\omega$ on $r$ ({\it e.g.}  $\zz$ evolution for the vacuum connection
(\ref{poframe})). Note that, analogously to the usual Hamiltonian
formalism,  to proceed one should first evaluate the derivatives
$\f{\p F(W_0,W_1)}{\p W_0^\Omega}$ over the general vacuum connection
 setting $W_0$ to its particular value like (\ref{poframe}) afterwards.
The second term in (\ref{ham})  describes an effective Hamiltonian in agreement
with \cite{act} where it was argued that  universal unfolded equations
(\ref{unf1}) provide a proper multidimensional generalization of the Hamiltonian
dynamics.

For the functional $S_\zz$ (\ref{Se}) analogous analysis gives
\be
\label{hamz}
\dot S_\zz(W_0,W_1) + H S_\zz(W_0,W_1)=\int_\Sigma L_\zz\,,
\ee
where $\Sigma$ is the boundary surface and $L_\zz$ is the component of $L$ along the
(real) $\zz$-direction,
\ie
$$L = d\zz L_\zz +\ldots\,.$$

An interesting aspect of the HS holography is the holographic interpretation of the
 RG flow  in terms of the bulk dynamical equations with respect to the radial coordinate
\cite{Douglas:2010rc,Sachs:2013pca,Leigh:2014qca,Mintun:2014gua}.
In  approach (\ref{S}) the RG-like equation controls  independence of
$S$  of the integration contour $S^1$. This interpretation is
reminiscent of the Wilsonian  approach  based on the independence of the
scale of fields distinguishing between UV and IR regions which
can be regarded as those inside and outside  $S^1$, respectively.
The analogy of Eqs.~(\ref{ham}) and (\ref{hamz}) with
the Hamilton-Jacoby approach of \cite{de Boer:1999xf} is also encouraging.
It would be interesting
 to check more closely the relation of Eq.~(\ref{ham}) to the holographic interpretation of
the RG flow in the boundary duals of the HS theories.

\section{HS equations in $AdS_4$}
\label{Nonlinear Higher-Spin Equations}

\subsection{Original nonlinear system}
In this section we recall the formulation of the $AdS_4$ HS field equations with the
emphasis  on their properties relevant to the construction of invariant
functionals.

 HS dynamics was formulated
in  \cite{more} in terms of zero-form $ B(Z;Y;\K|x)$,
space-time connection one-form $W(Z;Y;\K|x)$ and one-form connection $S(Z;Y;\K|x) $
in the $Z$-space. $W(Z;Y;\K|x)$ and $S(Z;Y;\K|x) $ can be combined into
the total connection one-form
\be
\W = \dr_x+ \theta^\un W_\un (Z;Y;\K|x)+ \theta^A S_A (Z;Y;\K|x)\q \dr_x =
\theta^\un \f{\p}{\p x^\un}\,,
\ee
where all differentials $dZ^A$ and $dx^\un$, denoted in the sequel
$\theta^A$ and $\theta^\un$, respectively, are anticommuting.
$A=1,\ldots 4$ and $\un =0,\ldots 3$ are indices of $4d$ Majorana spinors and
vectors, respectively. $Z^A$  and $Y^A$ are commuting spinorial variables.
Every Majorana spinor can be represented as a pair of two-component spinors with
$A=(\ga\,,\dga)$,
e.g., $\theta^A=(\theta^\ga, \bar \theta^\dga)$.
 $\K=(k,\bar{k})$ denotes a pair of Klein operators
that reflect  two-component spinor indices as
\bee
\label{kk}
k* w^\ga = -w^\ga* k\,,\quad &&
k *\bar w^\pa = \bar w^\pa *k\,,\quad
\bar k *w^\ga = w^\ga *\bar k\,,\quad
\bar k *\bar w^\pa = -\bar w^\pa *\bar k\,,\nn\\
&&\quad k*k=\bar k*\bar k = 1\,,\quad
k*\bar k = \bar k *k\,
\eee
 with
 $w^\ga= (y^\ga, z^\ga, \theta^\ga )$, $\bar w^\pa =
(\bar y^\pa, \bar z^\pa, \bar \theta^\pa )$. Note that relations (\ref{kk})
provide the definition of the star product with $k$ and $\bar k$.

The nonlinear HS equations of \cite{more} are
\be
\label{SS}
\W*\W= -i\left (\theta_A \theta^A + \delta^2(\theta_z)  F_*(B)* k*\ups +
\delta^2(\bar \theta_{\bar z}) \bar F_*(B) *\bar k *\bu\right )
\,,
\ee
\be
\label{SB}
\W*B=B*\W\,,
\ee
where
\be
\delta^2(\theta_z) = \half \theta_\ga \theta^\ga\q
\delta^2(\bar \theta_{\bar z}) = \half \bar \theta_\dga \bar \theta^\dga\,
\ee
and $F_*(B) $ is some star-product function of the zero-form $B$
\be
\label{F}
F_*(B) = \sum_{n=1}^\infty f_n \underbrace{B* B* \ldots * B}_n\,.
\ee

The simplest case of linear $F_*(B)$
\be
\label{etaB}
F_*(B)=\eta B \q \bar F_* (B) = \bar\eta B\,,
\ee
where $\eta=\exp [i\varphi]$, $\varphi\in [0,\pi)$ (the absolute value of $\eta$
can be absorbed into $B$) leads to a class of pairwise nonequivalent nonlinear HS
theories. The cases
of $\eta=1$ and $\eta =\exp{\f{i\pi}{2}}$ are particularly interesting, corresponding
to the so called $A$ and $B$  models that respect parity \cite{Sezgin:2003pt}.

The associative \emph{HS star product} $*$ acts on functions of two
spinor variables $Z_A$ and $Y_A$
\be
\label{star2}
(f*g)(Z;Y)=\frac{1}{(2\pi)^{4}}
\int d^{4} U\,d^{4} V \exp{[iU^A V^B C_{AB}]}\, f(Z+U;Y+U)
g(Z-V;Y+V) \,,
\ee
where
$C_{AB}=(\epsilon_{\ga\gb}, \bar \epsilon_{\dga\dgb})$
is the $4d$ charge conjugation matrix allowing to raise and lower indices
\be
Y^A = C^{AB}Y_B\q Y_A= Y^B C_{BA}\,,
\ee
and
$ U^A $, $ V^B $ are real integration variables. It is
normalized so that $1$ is the unit element, \ie $f*1 = 1*f =f\,.$
Star product (\ref{star2}) yields a particular
realization of the Weyl algebra
\be
\label{zzyy}
[Y_A,Y_B]_*=-[Z_A,Z_B ]_*=2iC_{AB}\,,\qquad
[Y_A,Z_B]_*=0\q[a,b]_*=a*b-b*a\,
\ee
and possesses a supertrace operation
\be
\label{str}
str (f(Z,Y)) = \frac{1}{(2\pi)^{4}}
\int d^{4} U\,d^{4} V \exp{[-iU^A V^B C_{AB}]}\, f(U;V) \,
\ee
respecting the cyclic property
\be
\label{cycl}
str(f*g ) = str (g*f)
\ee
provided that, in accordance with the normal spin-statistic relation, the
coefficients of the expansions of $f(Z;Y)$
in powers of spinor variables  $Z$ and $Y$ are (anti)commuting for  $f(Z;Y)$
(odd)even with respect to $f(-Z,-Y)= (-1)^{\pi_f}f(Z,Y)$
(and similarly for $g(Z;Y)$).

Star product (\ref{star2}) admits the inner Klein operator
\be
\label{ups}
\Upsilon = \exp i Z_A Y^A \,,
\ee
which obeys
\renewcommand{\U}{\Upsilon}
\be
\label{[UF]}
\U *f(Z;Y)=f(-Z;-Y)*\U\q \U *\U =1\,.
\ee
The left and right inner Klein operators
\be
\label{kk4}
\ups =\exp i z_\ga y^\ga\,,\qquad
\bu =\exp i \bar{z}_\dga \bar{y}^\dga\,,
\ee
 which enter Eq.~(\ref{SS}), act analogously
on  undotted and dotted spinors, respectively,
\be
\label{[uf]}
\ups *f(z,\bar{z};y,\bar{y})=f(-z,\bar{z};-y,\bar{y})*\ups\,,\quad
\bu *f(z,\bar{z};y,\bar{y})=f(z,-\bar{z};y,-\bar{y})*\bu\,,
\ee
\be
\ups *\ups =\bu *\bu =1\q \ups *\bu = \bu*\ups\,.
\ee

{}From (\ref{str}) and (\ref{ups}) it follows that
the supertrace of the inner Klein operators $\ups$ and $\bu$
diverges as $\delta^4(0)$  (for more detail
see \cite{Vasiliev:2015wma}). Hence, one has to be careful
with the expressions
defined as $str (f)$ for $f$ containing exponentials  behaving like $\ups$ and/or $\bu$.
This fact is of key importance for the further analysis since, as explained in Section \ref{struc},
 nontrivial invariant functionals considered in this paper should
 have divergent supertrace. From this perspective our approach is  opposite
to the construction of invariants in
\cite{Sezgin:2005pv,Sezgin:2011hq,Iazeolla:2011cb,Colombo:2012jx} where divergent supertraces were somehow
regularized. (Consistency  of such a regularization is not quite obvious
to us since the  star-product algebra admits a uniquely defined
supertrace.)

Naively,  field equations (\ref{SS}) and (\ref{SB}) leave no  room
for  a nontrivial invariant action written as a space-time  differential form built from
$\W$ and $B$. Indeed, since all space-time curvature tensors $\W* \W$ are zero by virtue
of the field equations as  well  as the star-commutator
$[\W\,,\B]_*$, $p$-form Lagrangians with $p>1$
like, e.g.,  $str(\W *f(B)*\W *g(B))$ are zero. One-form functionals $str(\W *f(B))$
are not gauge invariant. The zero-forms $str(f(B))$ at a point $x=x_0$ are the
invariants considered
 in \cite{Sezgin:2005pv,Colombo:2012jx}. These however are not well-defined due to
 divergencies of the supertrace. The trick explained in  Section \ref{struc} is to
 consider densities which, not being of  the form $str(L)$, would be exact if the trace
 operation was well defined but become nontrivial just because the respective trace is
 divergent.

As shown in \cite{Vasiliev:2015wma} perturbative analysis of the HS field equations leads
to solutions valued in the HS field algebra $\Sp$. General elements of $\Sp$ have divergent
supertrace. On the other hand, $\Sp$ contains a subalgebra $\Spl_0$ all elements of which
have finite supertrace. Therefore, to be nontrivial, $\Ll$ should be projected from
$\Sp/\Spl_0$. This observation will give us hints in Section \ref{invfunct}
on the structure of the system generating  the invariant density forms.

\subsection{Fock behavior at infinity}
\label{Fock}

In \cite{Vasiliev:2012vf} it was shown that the dependence  of the HS zero-forms
on the Poincar\'e coordinate
$\zz$ of  $AdS_4$ is
\be
\label{ct}
C(y,\bar y;\K|\xx,\zz) = \zz \exp  (y_\ga \bar y^\ga )T(w,
\bar w;\K|\xx,\zz)\,,
\ee
where $\xx$ and $\zz$ are, respectively, the boundary and Poincar\'e coordinates,
\be
\label{w}
w^\ga = \zz^{1/2} y^\ga \q \bar w^\ga =
\zz^{1/2} \bar y^\ga\,,
\ee
and $T(w,\bar w|\xx,\zz)$ is holomorphic in $\zz$.
The exponential factor in (\ref{ct}) leads to the nonpolynomiality of
the star-product element in the boundary limit of the HS theory. This
can lead to infinities in the local conformal limit $\zz\to 0$
as we discuss now.

As observed in \cite{Vasiliev:2012vf}, the exponential
\be
\label{fock}
F=4 \exp  y_\ga \bar y^{\ga}
\ee
provides the  star-product realization of the Fock vacuum that
satisfies
\be
y^-_\ga * F =F*y^+_\ga =0\,,
\ee
where
\be
y^+_\ga = \half (y_\ga - i \bar y_\ga)\q
y^-_\ga =\half (\bar y_\ga -i y_\ga)
\ee
obey
\be
\label{concom}
[ y^-_\ga\,, y^{+\gb }]_* = \delta_{\ga}^{\gb}\q [y^-_\ga\,,y^-_\gb]_* = 0\q
[y^{+\ga}\,,y^{+\gb}]_*=0\,.
\ee
$F$ is a projector, \ie
\be
\label{FF}
F* F = F\,.
\ee

The Klein operators $k$ and $\bar k$ exchange  $y^+ $ and $y^-$
\be
k y^\pm_\ga  = {\mp} i y^\mp_\ga  k\q
\bar k y^\pm_\ga = {\pm} i y^\mp_\ga  \bar k\,.
\ee
This implies that
\be
\label{bfock}
 \bar F = k F k =  \bar k F \bar k   =4 \exp  - y_\ga \bar y^{\ga}
\ee
obeys
\be
y^+_\ga * \bar F =\bar F*y^-_\ga =0\,.
\ee
$\bar F$ is also a projector\,,
\be
\label{bFbF}
\bar F* \bar F = \bar F\,.
\ee
However, the star product of $F$ with $\bar F$ is ill defined, being infinite
\be
\label{FFinfty}
F* \bar F = \infty\,.
\ee
This fact is  insensitive to the particular form of the star
product. Indeed, the relation
\be
F* \bar F =\f{1}{4} F* [ y^-_\ga\,, y^{+\ga }]_* * \bar F = 0 F* \bar F\,
\ee
demands $F* \bar F$ be either zero or infinity.
It is infinity for bosonic oscillators and zero for fermionic (if introduced).

Let $F(t)$  interpolate between $F$ and $\bar F$
\be
F(t) =4 \exp t y_\ga \bar y^{\ga}\q F(1)=F\q F(-1)= \bar F \,.
\ee
In accordance with \cite{Iazeolla:2011cb}, direct computation gives
\be
\label{Ftt}
F(t)* F(t') =  \f{4}{(1+t t')^2}\exp \left [ \f{t+t'}{1+t\,t'} y_\ga \bar y ^\ga
\right ]\,.
\ee
Eqs.~ (\ref{FF}), (\ref{bFbF}) and (\ref{FFinfty}) are particular cases of this formula.

Note that  formula (\ref{Ftt}) was used in \cite{Iazeolla:2011cb}
for the analysis of the HS BH solutions which  turns
out to be analogous to the  $AdS/CFT$ problem being
 based  \cite{Didenko:2009td} on the Fock vacuum  analogous to $F$
(\ref{fock}).
This analogy is very intriguing and suggestive.

HS equations (\ref{SS})  contain an arbitrary
star-product function $F_*(B)$ (\ref{F}) introduced in \cite{more}.
In the linearized approximation, the physical component of the field $B$
is given by (\ref{ct}) with $T(w,
\bar w;\K|\xx,\zz)$ proportional to either  $k$ or  $\bar k$. This implies that
 the product $B* B$ contains
\be
F* k * F = F* \bar F k =\infty
\ee
and similarly in the $\bar k$ sector. As a result, such terms exhibit infinities
 in the conformal limit where the behavior (\ref{ct}) is
imperative.\footnote{The authors of \cite{Iazeolla:2011cb} used such a regularization
of their computation scheme that $F*\bar F=0$. We do not quite see however how the
whole setting should be redefined to make this compatible with the associativity of a
sufficiently rich class of functions appropriate for the description of fluctuations.}
There are several ways for resolution of this problem. The simplest
 is to set  the non-linear terms in $F_*(B)$ in (\ref{F}) to zero. However, a
 HS theory with nonlinear $F_*(B)$ still may make  sense both in the bulk and
 at the boundary.

The point is that the two factors in the star products like
 $B(Z;Y;\K|x)*B(Z;Y;\K|x)$ are taken at the same $x$. This suggests that at the boundary
 they would correspond to operator products at the same $\xx$. The latter has to be
  regularized via an $\xx$-space point splitting. In the lowest order, the simplest way to
 achieve this is  by using formulae from Section 8 of \cite{Gelfond:2013xt} where they were
  used to reconstruct space-time current algebra from that in the twistor-like variables.
As will be shown in more detail elsewhere, the resulting expressions lead to finite results
with  split points of the product factors in the nonlinear terms of  $F_*(B)$.
  This suggests that seemingly local nonlinear terms  in $F_*(B)$  represent
  certain nonlocal terms in the boundary correlators.

While, naively, the terms   $B(x)* B(x)$  diverge in the boundary limit,
the terms linear in $B$ are free of such a divergency. A less trivial question
is whether the  terms linear in $B$ on the {\it r.h.s.} of the
nonlinear equations give rise to terms that remain finite in the boundary limit
 in the higher orders of the perturbative expansion.
Analysis of this question requires systematic investigation in spirit of
\cite{Vasiliev:2015wma} which has not been yet accomplished. Instead,
 we give here a simple indication that this has a chance to be true.

From formulae (\ref{w110}) and (\ref{ct}) it follows that
the star product of the first-order contributions involves the star product
of the exponentials of the following type
\be
\int_0^1 \rho_1(t_1) dt_1  \int_0^1 dt_2 \rho_2(t_2) \exp i[t_1 z_\ga (y^\ga +i\bar y^\dga)]*
\exp i [- t_2 z_\ga (y^\ga +i\bar y^\dga)]\,.
\ee
Evaluation of the star product yields the following integration measure
in $t_{1,2}$
\be
\int_0^1 \rho_1(t_1) dt_1 \int_0^1 \rho_2(t_2)  dt_2 \f{(1-t_1)(1-t_2)}{(1-t_1 t_2)^2}
\ee
which converges since
\be
 \f{1-t_1}{1-t_1 t_2}\leq 1 \q  \f{1-t_2}{1-t_1 t_2}\leq 1\q t_1,t_2\in [0,1]\,.
\ee
This  indicates that the contribution of the
terms linear in $B$ on the {\it r.h.s.} of (\ref{SS}) should make sense
in  the higher orders though the full analysis of this issue remains to be
done.

An interesting question for the future analysis of the boundary dual of
HS theory is to understand the difference between
conformal models resulting from linear and nonlinear $F_*(B)$.
One option is that the boundary conformal models associated with nonlinear
terms in $F_*(B)$  may themselves be nonlocal.

\section{Invariant density extension}
\label{struc}
All available HS equations like (\ref{SS}), (\ref{SB}) are particular examples of
the following system
\be
\label{hss}
\W *\W = \F_\Ll(c, \B,\Ll)\q \W*\B = \B*\W\q \dr_x\Ll =0 \,,
\ee
where $\W$ and $\B$  describe, respectively, forms of odd and even degrees both
in the space-time differentials $\theta_x$ and in the twistor-like  differentials $\theta_Z$,
whatever they are. More precisely
$\W=\dr_x +\W'$  contains the space-time de Rham derivative $\dr_x$ so that
the covariant derivative is equivalent to the commutator with $\W$.

Known HS theories are either directly based on some  associative algebra $A$,
where all fields are valued, or are  reductions of such theories. This allows an extension
of the system with fields valued in  the tensor
product of the original associative HS algebra $A$ with any internal associative
algebra $A_{int}$ bringing  Chan-Paton-like indices carried by all fields.
Note that supersymmetric HS theories result from
this construction with $A_{int}$ being a Clifford algebra \cite{V3,KV}
(see \cite{Vasiliev:1999ba,Bekaert:2005vh} for more detail and
\cite{Chang:2012kt} for recent applications).
In the sequel, the  internal indices will be implicit with the convention that
central elements of   HS algebras are valued in the center of $A_{int}$,
that is in  the unit matrix of $A_{int}=Mat_n(\mathbb{C})$.

We introduce the following conventions.
The  $*$ in (\ref{hss}) is the product in $A$. In the $AdS_4$ HS system,
$A$ is the star-product algebra (\ref{star2}) of functions $f(Z;Y;\K)$. The tensor product
of $A$ with the wedge algebra of differentials $\theta_x, \theta_Z$ will be denoted
$\Lambda A$. Analogously, $\Lambda_x A$ and $\Lambda_Z A$ denote the tensor products
of $A$ with the wedge algebra of differentials $\theta_x$ and  $\theta_Z$, respectively.
$C$, $\Lambda C$, $\Lambda_Z C$ and $\Lambda_x C$ are, respectively, the centers of
$A$, $\Lambda A$, $\Lambda_Z A$ and $\Lambda_x A$. Fields of the theory
are $x$-dependent differential forms valued in $\Lambda A$ (\ie sections of the respective
fiber bundles).

In (\ref{hss}), $c$ are $x$-independent elements of $\Lambda_Z C$.
Central elements $c$ which appear in the original $AdS_4$ HS theory
are  $I$, $\theta_A \theta^A$, $\delta^2(\theta)  k*\ups$ and
$\delta^2(\bar \theta) \bar k *\bu$. They play different r\'oles
in the theory.
For instance, $\theta_A \theta^A$ does not belong to the field
HS algebra $\Sp$ of \cite{Vasiliev:2015wma} where it was shown that it cannot appear anywhere
in the nonlinear field equations except for the first term on the
{\it r.h.s.}  of (\ref{SS}) if solutions of the HS equations are demanded to
be minimally nonlocal, belonging to $\Sp$. This is related to the fact that
the central element $-i\theta_A \theta^A$ is the square of the operator
\be
\label{Q}
Q=\theta^A Z_A
\ee
 which determines the $Z$-dependence of the HS fields via
perturbative solution of  system (\ref{SS}), (\ref{SB}) (for more detail
see Section \ref{sketch}) and which also does not belong to $\Sp$ \cite{Vasiliev:2015wma}.
Hence,  we will demand  the central elements that affect interactions
to belong to $\Lambda C_{\Sp}\subset  \Sp$.

New ingredients of the construction are
 space-time differential forms  $\Ll$  valued in $\Lambda_x C$. As explained in
 Section \ref{sketch} the forms $\Ll$  associated with central elements
$c_0\in C_\Sp$, that belong to the $Q$-cohomology in $\Sp$ for $Q$ (\ref{Q}),
play a distinguished role. In the twistorial HS theories considered in this paper
the $Q$-cohomology is represented by the
zero-forms in the $\theta_Z$-space, \ie by the unit element
of the star-product algebra. In this case $\Ll=\Ll(\theta_x|x)$ depends only on the
space-time coordinates $x$ and differentials $\theta_x$.

System (\ref{hss}) is consistent with respect to further
commutators with $\W$ (\ie covariant differentiation) because $\B$ commutes with itself,
as well as with all $c$ and $\Ll$. It is invariant under the following gauge transformations
with  three types of gauge parameters $\gvep$, $\xi$ and $\chi$:
\be
\label{dw}
\delta \W = [\W\,, \gvep]_*\,+  \xi^N \f{\p \F_\Ll(c,\B,\Ll)}{\p \B^N}+
\chi_i \f{\p \F_\Ll(c,\B,\Ll)}{\p \Ll_i}\,,
\ee
\be
\label{db}
\delta \B = \{\W\,,\xi\}_* + [\B\,, \gvep]_* \,,
\ee
 \be
 \label{dl}
\delta \Ll_i = \dr_x \chi_i\q
\ee
where $N$ is the  multiindex running over all components of $\B$,
the gauge parameters
$\gvep$ and $\xi$ are differential forms of  even and odd
degrees, respectively, being otherwise  arbitrary functions of coordinates
and the generating elements of the star-product algebra, while $\chi_i$ only depend
on the  space-time coordinates and differentials. For instance, in  the $AdS_4$
HS theory
\be
\gvep=\gvep(\theta; Z; Y;\K|x)\q \xi=\xi(\theta; Z; Y;\K|x)\q
\chi = \chi (\theta_x|x)\,.
\ee
Transformations (\ref{dw}) are usual HS gauge transformations
extended to higher differential forms.
Transformations (\ref{db}) are their analogues for
higher-form components of $\B$. (Such transformations were considered in
 \cite{Boulanger:2011dd}.) Gauge transformation (\ref{dl}) implies equivalence
of the forms $\Ll_i$ modulo exact forms. All three types of  gauge
transformations provide a realization of gauge transformations (\ref{delw})
for various differential forms in the system.

The  most important case of appearance of the
 forms $\Ll_i$ is additive with
\be
\label{add}
\F_{\Ll}(c, \B,\Ll)= \F(c, \B) + \Ll_i \,c_0^i\,,
\ee
where $c_0^i\in  C_\Sp$. For the models considered in this paper
the only option is  $c_0=I$.

In the additive case,   $\Ll_i$ are expressed by the first equation in (\ref{hss})
in terms of the other fields.
In this case $\dr_x\Ll_i =0$ is not an independent
condition but rather a consequence of the other equations in system (\ref{hss}).
Note that this may not be true for those  $c^i\in \Lambda \Sp$ that do not
belong to $Q$ cohomology because, containing a product with some nonzero power of $\theta_Z$,
in this case the
compatibility condition for  system (\ref{dw}) restricts the derivative of
$\Ll_i$ modulo terms that give zero upon multiplication by $c^i$.

The gauge transformation of
the $(p_{\Ll_i} -1)$-forms $\W_{\Ll_i}$ valued in the same central elements $c_0^i$
is
\be
\label{dwc}
\delta \W_{\Ll_i} = \pi_{i}\Big ([\W\,, \gvep]_*\,+  \xi^N \f{\p \F(c,\B)}{\p \B^N}\Big)
+
\chi_i\,,
\ee
where $\pi_{i}$ is the projection to the central element $c_0^i$. For example,
for the usual HS algebra, the projection to the unit element $I$ realized by
 $f(Z,Y)=1$ is
\be
\pi_I (f(Y,Z|x)) = f(0,0|x)\,.
\ee
Let us stress that for  additive  systems the gauge fields $\W_{\Ll_i}$
from the center of the algebra $\Sp$ become locally trivial being pure gauge
with respect to the $\chi$-transformation in (\ref{dwc}). Locally, one can therefore
use {\it canonical gauge}
\be
\label{cang}
\W_{\Ll_i}=0\,.
\ee

The {\it density forms} $L_i$ are
\be
\label{L}
L_i = \Ll_i- \dr_x \W_{\Ll_i}\,.
\ee
From (\ref{hss}), (\ref{dl}) and (\ref{dwc}) it follows that they are $\dr_x$-closed
\be
\label{dL}
\dr_x L_i=0\,
\ee
and their gauge transformation
\be
\label{gtl}
\delta L_i= - \dr_x \pi_{i}\Big ([\W\,, \gvep]_*\,+  \xi^N \f{\p \F(c,\B)}{\p \B^N}\Big)\,
\ee
 is $\chi$-independent which allows them to be
 globally defined. Note that  $\Ll_i$ and $L_i$
coincide in the canonical gauge (\ref{cang}).

The density forms $L_i$ can represent differential forms of different degrees $p$
\be
L_i = \sum_p L_i^p\,.
\ee
Functionals defined as integrals of the density forms over closed $p$-cycles $\Sigma^p$
 in the $x$-space
\be
S^p_i =\int_{\Sigma^p} L^p_i\,
\ee
 are gauge invariant because the density
forms transform by a total derivative (\ref{gtl}) under the gauge transformations of
the other fields in the system.
Whether the resulting invariants are non-zero or trivial depends on
a particular solution of the theory. Since the forms $L^p_i$ are closed, the
result of their integration  can be non-zero only for
non-contractible cycles, \ie for singular solutions. As explained in Introduction,
in  the case of $AdS/CFT$ the singularity is at infinity and $L^4$ is a four-form
in the complexified $AdS_4$ case. For $AdS_4$ BH solutions of
\cite{Didenko:2009td,Iazeolla:2011cb,Bourdier:2014lya}
the corresponding invariants are supported by the two-forms $L^2$. As argued in Section
\ref{Black holes}, these should reproduce
the BH charges in the gauge invariant way with the invariant functional $S$
saturated by the BH singularity.

Suppose now that the HS algebra $A$ possesses a supertrace
obeying the cyclic property (\ref{cycl}). Let $c^*_i$ be the central elements
dual to $c^i$ in the sense that
\be
\label{strcc}
str (c^i * c^*_j) = \delta^i_j\,.
\ee
For instance, for $c=I$ it is convenient to normalize the supertrace so that
$str (I)=1$ and $I^* = I$.
Setting the differentials $\theta_Z$  in $F(c, \B)$  to zero yields
\be
\label{lstr}
L_i = str \big ( c^*_i *\W'* \W'\big )\Big |_{\theta_Z=0}\,.
\ee
Since $\W'$ are forms of odd degrees, from the cyclic  property (\ref{cycl})
 it seemingly follows that  $L_i=0$.
There is a subtlety however that  formula (\ref{lstr}) can be ill defined if
$str (c^*_i*(\W'* \W'))$
is divergent. In other words, as explained in \cite{Vasiliev:2015wma}, the actual class of
functions valued in $\Sp$ that appear in the analysis is wider than the class of functions
admitting the supertrace which form the algebra  $\Spl_0$ \cite{Vasiliev:2015wma}. We will
see an example of this phenomenon in Section \ref{Black holes}.

Alternatively, it can happen that some of central elements $c$ admit no $c^*$
obeying (\ref{strcc}). For instance, $c^*$ does not exist
if $str (I)=0$ which case is known to play a role in the maximally
supersymmetric  $N=4$ SYM.
This gives a criterion distinguishing between
trivial and non-trivial densities: those $L_i $, for which
$str (c^*_i*(\W'* \W'))$ is well defined, are trivial while those for which it is
ill defined either being divergent or because $c^*_i$ obeying (\ref{strcc})
does not exist have a chance to be nontrivial.

The construction of this section  exhibits  essential difference between
densities of even and odd degrees. Indeed, in the additive case,
$\Ll$  appears on the {\it r.h.s.} of the first of  equations (\ref{hss})
whose {\it l.h.s.} is the square of odd forms $\W$. Hence, the density form associated
with any central element, which is even in the differentials
as is the case in all known examples, is a form of some even degree
while the density forms associated with central
elements of degree zero like unit element $I$ must have strictly
positive even degree.

In the sequel of this section
we consider a  more general construction which can also lead
to density forms of odd degrees. To this end system (\ref{hss}) can be
 modified  to
\be
\label{hssg1}
\W *\W = \F_\Ll(c, \B,\Ll)\,,
\ee
\be
\label{hssg2}
\W * \B - \B* \W = \Gg_\Ll(c, \B,\Ll)\,,
\ee
\be
\label{hssg3}
\dr_x\Ll =0 \,,
\ee
where $\F_\Ll$ and  $\Gg_\Ll$ are even and odd differential forms, respectively.
In the  case of $\Gg_\Ll=0$ we recover system (\ref{hss}) free of any
restrictions on $\F_\Ll$.
Compatibility of (\ref{hssg1}), (\ref{hssg2}) demands
\be
\label{comp}
\W* \F_\Ll(c,\B,\Ll) - \F_\Ll(c,\B,\Ll) *\W =0\q
\W* \Gg_\Ll(c,\B,\Ll) + \Gg_\Ll(c,\B,\Ll) *\W =0\,
\ee
and, hence,
\be
\label{com}
\Gg^N_\Ll(c, \B,\Ll) \f{\p \F_\Ll(c, \B,\Ll)}{\p \B^N} =0\q
\Gg^N_\Ll(c, \B,\Ll) \f{\p \Gg^M_\Ll(c, \B,\Ll)}{\p \B^N} =0\,.
\ee

System (\ref{hssg1})-(\ref{hssg3}) is invariant under   gauge transformations
(\ref{dw}) and (\ref{dl}) for $\W$ and $\Ll$
while the transformation law for $\B$ modifies to
\be
\label{dbg}
\delta \B = \{\W\,,\xi\} + [\B\,, \gvep]_*
+  \xi^N \f{\p \Gg_\Ll(c,\B,\Ll)}{\p \B^N}+
\chi_i \f{\p \Gg_\Ll(c,\B,\Ll)}{\p \Ll_i}\,.
\ee
Main features of the analysis of the gauge transformations
remain the same as at $\Gg=0$. The novelty is that the components of the fields $\B_\Ll$
associated with the forms $\Ll$ in $\Gg_\Ll$ on the {\it r.h.s.}
of (\ref{hssg2}) become pure gauge.

Conditions (\ref{com}) have the following interesting interpretation.
The second condition implies that the odd vector field
\be
\label{qg}
\Q := \Gg^N_\Ll \f{\p}{\p \B^N}
\ee
is nilpotent
\be
\label{qq}
\Q^2 =0\,.
\ee
The first  implies that $\F_\Ll$ must be $\Q$-closed
\be
\label{qf}
\Q \F_\Ll =0\,.
\ee
$\Q$-exact $\F_\Ll$
\be
\F_\Ll(c,\B,\Ll) = \Q f(c,\B,\Ll)
\ee
are dynamically trivial since they can be removed by a field redefinition
of $\W$ which in the infinitesimal case is
\be
\delta \W = f(c,\B,\Ll)\,.
\ee

Thus,  general HS system (\ref{hssg1})-(\ref{hssg3}) is characterized by a
nilpotent vector field $\Q$ (\ref{qg}) and some its cohomology $\F_\Ll$.
Similarity of this construction with the description
of unfolded systems in Section \ref{fda} is obvious. Note however that
 systems  with $\Q=0$ are  nontrivial and,
in fact, most interesting while unfolded equations (\ref{unf1})
with $Q=0$ are trivial.

 The general case with $\Gg_\Ll\neq 0$  may also have applications.
Let us note however that, to fulfill  conditions (\ref{com}),
$\Gg_\Ll$ should contain such a combination of the $\theta_Z$ differentials
that it would give zero upon multiplication with the $\B$-derivatives
of $\F_\Ll(c,\B,\Ll)$ and $\Gg_\Ll(c,\B,\Ll)$ in (\ref{com}). This is
impossible for the forms $\Ll$ associated with central elements $c_0^i\in \Sp_0$
which have zero degree in the differentials $\theta_Z$. In particular, for the
twistorial $3d$ and $4d$ models considered in this paper, where the HS field
algebra $\Sp$ is known \cite{Vasiliev:2015wma}, we were not able to construct forms $\Ll$
of odd degrees. However, for more general models like vectorial
HS models of \cite{Vasiliev:2003ev} a proper generalization of the  construction
of \cite{Vasiliev:2015wma}  of the HS field algebra remains unknown.
For this  case not only the structure of the HS field algebra can be changed
but also the structure of cohomology of the respective generalization of $Q$ (\ref{Q})
(not to be confused with $Q$ (\ref{qdif})).
If cohomology  $H^p(Q)$  with $p>0$ has  nonzero
components in the center of the HS algebra,
 the construction of this section can lead to
nontrivial invariants. It would be interesting  to apply it
to  the vectorial HS models of \cite{Vasiliev:2003ev}.

\section{Invariants of $AdS_4$ HS theory}
\label{invfunct}
\subsection{Extended system}

To construct invariant functionals of the $AdS_4$ HS theory we
 extend  system (\ref{SS}), (\ref{SB}) as follows.
 $\W(\theta;Z;Y;\K|x)$ is extended to all odd forms
 while $B(Z;Y;\K|x)$ is extended to all even forms $\B(\theta;Z;Y;\K|x)$, \ie
 $\W(\theta;Z;Y;\K|x)$ is a polynomial of $\theta_x$ and $\theta_Z$
 of total degrees $1,3,5,\ldots$, while $\B(\theta;Z;Y;\K|x)$ is a polynomial of
 $\theta_x$ and $\theta_Z$ of total degrees $0,2,\ldots$.
(Such an extension was considered, {\it e.g.}, in \cite{333,Boulanger:2011dd}.)
Also we introduce the  forms of even degrees
\be
\label{L24}
\Ll(\theta_x|x)=\Ll^2(\theta_x|x)+\Ll^4(\theta_x|x)+\ldots \,,
\ee
 that only depend on the space-time coordinates and differentials.

 The extended HS system has the form (\ref{hss}) with
 \be
 \label{leq}
\ls F_\Ll(c, \B,\Ll) =-i \Big (\theta_A \theta^A + \delta^2(\theta_z)  F_*(\B)* k*\ups +
\delta^2(\bar \theta_{\bar z}) \bar F_*(\B) *\bar k *\bu +\delta^4(\theta_Z) G_*(\B)*
k*\bar k * \ups  *\bu + \Ll (\theta_x|x)\Big ) \,,
 \ee
 \be
\label{G}
 G_*(\B) = g +g_1 \B +g_2 \B*\B \ldots\,.
 \ee
 The overall factor of
 $-i$ in (\ref{leq}) is introduced to have real $ G_*(\B) $ and $\Ll$ with
 anti-Hermititian  $\W$ in
 (\ref{hss}). The simplest  case of
 $G_*(\B) = g=const$ is, in fact,  most interesting.

The  extended system is chosen in this form because the additional terms should belong to the
HS field algebra $\Sp$ introduced in \cite{Vasiliev:2015wma}
where it was shown that the central elements
$\delta^2(\theta_z)  * k*\ups$ and $\delta^2(\bar \theta_{\bar z})  *\bar k *\bu$
do belong to $\Sp$  while
$\delta^2(\theta_z) $, $\delta^2(\bar \theta_{\bar z}) $  and
$\delta^4(\theta_Z)$ do not. This means that, surprisingly, being in a certain
sense singular, the
latter operators  are not allowed to appear in the  HS system
with the only exception for the first term in (\ref{leq}) compensating the
``singularity" of  $Q$ (\ref{Q}) yielding an exterior derivation
in $\Sp$ \cite{Vasiliev:2015wma}.

Thus all $\B$-dependent terms in (\ref{leq}) belong to $\Sp$.
The perturbative analysis of Section \ref{sketch} shows that the
$g$-depended four-form in the twistor space induces invariant density
 forms of degrees  four and higher. According to the
analysis of Section \ref{Nonlinear Higher-Spin Equations}, the presence
of the Klein operators in the $G$-term of (\ref{leq}) gives rise to divergent traces
and, hence,  nontrivial densities.

The density form $L^4$ is anticipated to give rise to
 the generating functional of correlators in the $AdS_4/CFT_3$ HS holography.
Since the expression for  $L^4$ in terms of dynamical fields turns out to be
proportional to $g$, the latter acquires the meaning of the (inverse) coupling
constant in front of the Lagrangian ($N$ within the $1/N$ expansion)  also
 containing the inverse Planck constant $\hbar^{-1}$ in the  generating functional for
boundary correlators. The absence of such a constant  in original system (\ref{SS}),
(\ref{SB}) complicated its holographic interpretation.
Extended system (\ref{hss}), (\ref{leq}) contains the missed
elements  appropriate for the description of the quantum regime of the boundary theory.
Note that,  to account higher quantum corrections, it may be necessary
to consider higher-order differential forms in $\W$, $\B$ and $\Ll$ contributing to
higher-order corrections in $g$ via terms integrated over
Cartesian products of the original space-time with multiple space-time integrations
mimicking  loop integrations.

As explained  in Section \ref{Black holes}, the density two-form $L^2$
  supports the BH charges in the $4d$ HS theory. This should be
saturated by nontrivial BH-like solutions
\cite{Didenko:2009td,Iazeolla:2011cb,Bourdier:2014lya}
 of the original HS system (\ref{SS}), (\ref{SB}) with the BH mass
being a counterpart of $g$ via the contribution of a BH solution to the {\it r.h.s.}
of (\ref{SS}).

\subsection{Vacuum solution}

As  usual, we consider a vacuum solution with $\B=0$.
In the one-form sector it has the form
\be
\label{vac1}
\W_0 =\dr_x + \W_0^{1,0} + \W_0^{0,1} \q\W_0^{1,0} = Q\q \W_0^{0,1} = W_0(Y|x)\,,
\ee
where  the space-time one-form $W_0(Y|x)$ (the differentials $\theta_x$ are
implicit) is some solution to the flatness equation
\be
\label{W0eq}
\qquad\dr_x W_0(Y|x) +W_0(Y|x)*W_0(Y|x) =0\,.
\ee
For bilinear $W_0(Y|x)$
\be
\label{W0}
W_0(Y|x) = \f{i}{4}\, W_0^{AB}(x) Y_A Y_B\,
\ee
 (\ref{W0eq}) implies that the components  $W_0^{AB}(x)$ describe locally
$AdS_4$ geometry provided that the frame one-form $e^{\ga\dga}(x):=W_0^{\ga\dga}(x)$
is nondegenerate.

By virtue of (\ref{zzyy}), the star-commutator with
$\W_0^{1,0} = Q$ (\ref{Q}) is proportional to the de Rham derivative
in $Z^A$
\be
\label{dzf}
Q * f(Z;Y) -(-1)^{deg_f} f(Z;Y)*Q = -2i \dr_Z f(Z;Y)\q
\dr_Z = \theta^A \f{\p}{\p Z^A}\,
\ee
where $deg_f$ is the form degree of $f$. We use notation
\be
\label{wpq}
\W = \sum_{p,q} \W^{p,q}\,,
\ee
and $\W^{p,q}$ is a $p$-form in
the $Z$-differentials $\theta_Z$ and a $q$-form in the $x$-differentials
$\theta_x$.

Clearly, Eq.~(\ref{vac1})  gives a solution to (\ref{leq}) at $g=0$.
For  $g\neq 0$
it suffices to find the deformation of (\ref{vac1}) linear in $g$
since higher-order terms in $g$ contribute to forms of degrees six or higher
irrelevant in this paper.  To this end one can use the
standard homotopy formula for the de Rham derivative which is easy to check by
differentiation
\be
\label{homm1} \dr_Z f(\theta_Z; Z; Y)=g( \theta_Z; Z; Y)
\quad \Longrightarrow\quad
f (\theta_Z; Z; Y)= \pp g  +\dr_Z\gvep + f(0;0;Y)\,,
\ee
where
\be
\label{pzp}
\pp g :=\dr^*_Z H(g)\q H(g):=
\int_0^1 dt t^{-1} g(t \theta_Z; tZ; Y)\q
 \dr_Z^* = Z^A\frac{\partial}{\partial \theta^{A}}
\,.
 \ee
The term  $\dr_Z\gvep $ in Eq.~(\ref{homm1})
describes the freedom in exact forms while $f(0;0;Y)$ represents
the  de Rham cohomology.
 Eq.~(\ref{homm1}) is valid provided that the homotopy integral over $t$ converges,
which, in accordance with the Poincar\'e lemma, is true if  $g(0;0;Y)=0$. Note that
\be
\label{pzpz}
\pp \pp =0
\ee
since
\be
\dr_Z^*\dr_Z^*=0\,.
\ee

Equipped with these formulae it is straightforward to obtain the $(M-1)$-form
components $\W_0^{p,q}$ for the general case of  $A=1,\ldots ,M$. The final
result has the concise form
\be
\label{w03}
\W_0^{M-1} = \f{g}{2} Z^A \f{\p}{\p \theta^A} \int_0^1 d\tau \tau^{M-1}
\exp i\Big [\tau Z_A Y^A
+(1-\tau)W_0^{AB}(x) Z_A \f{\p}{\p \theta^B}\Big ]\delta^M(\theta ) k\bar k\,.
\ee
For $M=4$ this yields
\be
\label{vacs}
\W_0^{3-q,q} = \f{ g}{2}\sum_{q=0}^3\int_0^1 d\tau \tau^3\f{i^q(1-\tau)^q }{q!}
\exp{[i\tau Z_A Y^A]}\, Z^B \W_{0}^{A_1} (Z)\ldots \W_{0}^{A_q} (Z)
\delta_B{}_{A_q\ldots A_1}(\theta_Z)k\bar k
\,,
\ee
where
\be
\W_{0}^{B} (Z|x)=W_0^{AB}(x) Z_A \q
\delta^M_{A_1\ldots A_q}(\theta_Z) = \f{\p}{\p \theta^{A_1}}\ldots \f{\p}{\p\theta^{A_q}}
\delta^M(\theta_Z)\,.
\ee

An important property of  $\W_0^{0,3}$, which has to be checked separately to make sure that
it obeys   (\ref{leq}) with $\B=0$, is that
\be
\label{cow}
\dr_x \W_0^{0,3}(Z;Y;\K|x) +W_0(Y|x)* \W_0^{0,3}(Z;Y;\K|x) +\W_0^{0,3}(Z;Y;\K|x) * W_0(Y|x)  =0\,.
\ee
This follows from the observation that the star-commutator of the
{\it l.h.s.} of (\ref{cow}) with $\W_0^{1,0}=\theta^A Z_A$ is zero as a consequence of the
other vacuum equations which
have been already resolved, leading to (\ref{vacs}). On the other hand, the substitution
of (\ref{vacs}) into (\ref{cow}) gives  terms that are  zero at $Z=0$.
Hence the {\it l.h.s.} of ({\ref{cow})  is zero for any $Z$.
Straightforward verification of (\ref{cow}) involves a partial integration over $\tau$.

\subsection{Sketch of the first order}
\label{sketch}
Let
\be
\W = \W_0 +\W_1 +\ldots\q \B =\B_1 +\ldots\,,
\ee
where $\W_1$ and $\B_1$ are first-order fluctuations. The $\Ll$-independent part
of  linearized  equations (\ref{hss}) is
\be
\label{dW1}
{\rm d} \W_1 + \W_0 *\W_1 +\W_1*\W_0 =
-i \Big ( \eta\delta^2(\theta_z)  \B_1* k*\ups + \bar \eta
\delta^2(\bar \theta_{\bar z})  \B_1* \bar k *\bu \Big )\,,
\ee
\be
\label{dB1}
{\rm d} \B_1 +\W_0 *\B_1 -\B_1 *\W_0 =0\q \dr=\dr_Z+\dr_x\,.
\ee
Since $\W_0$ contains $\W_0^{1,0}$ proportional to ${\rm d_Z}$ (\ref{dzf}) these equations
express all components of $\W_1$ that are not ${\rm d_Z}$ closed via
 other fields. ${\rm d_Z}$-exact fields are pure gauge with respect to gauge
transformations (\ref{dw}), (\ref{db}). Hence, the remaining {\it physical
fields}, that are neither
expressed via the other fields nor pure gauge with respect to
the part of the gauge transformations containing  ${\rm d_Z}$, are in  the
${\rm d_Z}$-cohomology.

By Poincar\'e Lemma, these are fields independent of both $Z^A$ and
$\theta^A$, \ie
\be
C(\theta_x;Y;\K|x):= B_1(\theta;Z;Y;\K|x)\Big |_{\theta_Z=Z=0}
\ee
and
\be
\go(\theta_x;Y;\K|x):= \W_1(\theta;Z;Y;\K|x)\Big |_{\theta_Z=Z=0}\,.
\ee
$C(\theta_x;Y;\K|x)$ and $\go(\theta_x;Y;\K|x)$ contain
 space-time forms of even and odd degrees, respectively,
\be
\label{C03}
C(\theta_x;Y;\K|x)= C^0(Y;\K|x)+ C^2(\theta_x;Y;\K|x)+\ldots\,,
\ee
\be
\label{go}
\go(\theta_x;Y;\K|x)= \go^1(\theta_x;Y;\K|x)+ \go^3(\theta_x;Y;\K|x)+\ldots\,.
\ee
$C^0(Y;\K|x)$ and  $\go^1(Y;\K|x)$ are the HS fields of the original system.
 $C^p(Y;\K|x)$ and $\go^{p+1}(Y;\K|x)$ with even $p\geq 2$ are new.
 Note that most of  components of
 $C^p(Y;\K|x)$  are expressed via derivatives of
$\go^{p+1}(Y;\K|x)$  by  (\ref{dW1}).

The situation with densities is analogous: nontrivial densities should appear
in combination with those central elements $c^i_0$ in (\ref{add})
that belong to the $Q$-cohomology. Indeed, being central, $c^i$ is
$Q$-closed. If it is $Q$-exact, $c^i=[Q\,, \chi^i]_\pm$, in the lowest order, the term with
$\Ll$ can be removed by the  transformation
\be
\label{trl}
\W_1\to \W_1'=\W_1 -\chi^i \Ll_i \q \Ll_i\to 0\,,
\ee
which is a consequence of the following perturbative symmetry with the parameter
$\ga$
\be
\W_1\to \W_1'=\W_1 +\alpha \chi^i  \Ll_i \q  \Ll_i \to\Ll_i' =(1+\alpha) \Ll_i\,.
\ee
Hence, only central elements in the $Q$-cohomology $H(Q)$  generate
nontrivial densities.\footnote{I am grateful to Nikita Misuna for the
illuminating discussion of this point.}

For the  de Rham derivative $Q$ (\ref{Q}) acting on the freely generated
functions of $Z$  this implies by Poincar\'e lemma that nontrivial densities
are associated  with the unit element of the star-product algebra as in
(\ref{leq}). On the other hand, the terms
\be
\label{T}
 \delta^2(\theta_z)  * k*\ups\,\T(\theta_x|x) +
\delta^2(\bar \theta_{\bar z})  *\bar k *\bu \bar \,\overline \T(\theta_x|x)
\ee
with conjugated  $\T$ and $\overline \T$, that can also be added to the {\it r.h.s.}
of (\ref{leq}) provided that differential forms of higher degrees
 among $\W$ and $\B$ are introduced, unlikely give rise to nontrivial densities.
It would be instructive to understand the condition that $c^i$ should belong to
$H(Q)$ in more general terms either defining the actions in terms of
certain integrals over $Z$-variables to which $Q$-exact terms do not contribute or
 to trace the origin of  symmetry (\ref{trl}) back to extended symmetries considered
in Conclusion of \cite{Vasiliev:2015wma}.
In this paper  we just postulate that the densities are associated with the
central elements $c_0^i$ in $H(Q)$.

Straightforward  analysis of Eqs.~(\ref{dW1}), (\ref{dB1})
is technically involved, requiring more efficient tools explained in particular
in \cite{Didenko:2015cwv}. Here we only  mention some general aspects.

The one-form sector of (\ref{dB1}) gives
\be
\B_1^0(Z;Y;\K|x) = C^0(Y;\K|x)\,
\ee
and
\be
\label{dc}
{\rm d}_x C^0 (Y;\K|x)+W_0(Y|x) *C^0(Y;\K|x) - C^0(Y;\K|x)* W_0(Y|x) =0\,.
\ee According to the standard analysis of the
HS field equations  \cite{more,Vasiliev:1999ba}, $C^0(Y;\K|x)$
is the generating function for all
gauge invariant degrees of freedom in the system.
The fields $C_{\ga_1\ldots \ga_n}(x)$ considered in Introduction
are primary components of $C^0(Y;\K|x)$ in the conformal frame \cite{Vasiliev:2012vf}.

The two-form sector of (\ref{dW1}) gives
\be
\label{w110}
\W_1^{1,0} = \half \eta \int_0^1 dt \,t e^{itz_\ga y^\ga}z_\ga\theta_z^{\ga }
C(-tz,\bar y;\K)k + \half \bar \eta \int_0^1 d\bar t \,\bar t \exp^{i\bar t\bar z_\dga \bar
y^\dga}\bar z_\dga \bar \theta_{\bar z}^\dga
C(y,-\bar t \bar z;\K)\bar k
\ee
and
\be
\label{w101}
\W_1^{0,1} = -\f{i}{2} \p_Z \{W_0\,,\W_1^{1,0}\}\,,
\ee
where   the term with $\dr_x$ in $D_0$ does not contribute  because of (\ref{pzpz}).
Plugging (\ref{w101}) into the $\theta_x^2$ sector of (\ref{dW1}) yields
the so-called First On-Shell Theorem
\bee
\label{FOST}
&&\dr_x\go^1(Y;\K|x)+ W_0(Y)*\go^1(Y;\K|x) + \go^1(Y;\K|x)* W_0(Y)=\nn\\
&&=\f{i}{8}\Big (\eta \overline H^{\dga\dgb}
\f{\p^2}{\p \by^\dga \p \by^\dgb} C^0(0,\by;\K|x )k +\overline \eta
H^{\ga\gb}  \f{\p^2}{\p y^\ga \p y^\gb} C^0(y,0;\K|x )\bar k\Big )\,,
\eee
where
\be
\label{HH}
\overline H^{\dga\dgb}=
e^{\ga\dga}e_{\ga}{}^\dgb\q H^{\ga\gb}= e^{\ga\dga}e^{\gb}{}_\dga \q
e^{\ga\dga}:=W_0^{\ga\dga}\,.
\ee

First On-Shell Theorem imposes  spin $s>1$  equations on the
frame-like connections contained
in $\go^1(Y;\K|x)$. Eq. (\ref{dc}) contains the field equations
for spins $s\leq 1$. In addition, Eqs. (\ref{FOST}), (\ref{dc}) express  infinitely many
auxiliary fields via derivatives of the frame-like connections and matter fields
\cite{Ann,more} (see also \cite{Vasiliev:1999ba}).

To find  $\W_1$, which eventually determines the forms $\Ll$ in (\ref{leq})
one has to find  $\B_1$.
Reconstruction of these fields by the homotopy  formula (\ref{homm1}) is
straightforward but lengthy. Leaving details for \cite{DMV}, here
we would like to stress that the multiple application of the formula (\ref{homm1})
to the products of the
$g$-dependent part of the vacuum field (\ref{vacs})  with
the first-order HS fields $\go^1$ and $C^0$ reconstructs the first-order contributions
to  $\B_1$ and $\W_1$.
So, the contributions to the higher-form connections and, eventually, to
the invariant densities are induced by the $g$-dependent term in (\ref{leq}), (\ref{G}).
The bilinear part of $L^4$
\be
\pi_I\big(\{\W_1\,, \W_1\}_* +\{\W_2\,, \W_0\}_*\big)
\ee
 contains both $\W_1$ and  the second-order part $\W_2$ of $\W$
which needs another involved computation.  By this procedure
the quadratic part of $L^4$ turns out to be proportional to $g$.
The full density contains higher-order corrections  which can be
reconstructed order by order from  (\ref{hss}).

The only subtlety of this analysis is that,
 apart from  straightforward application of  homotopy
formula (\ref{homm1}), to reconstruct all perturbations one has to solve
the seemingly differential equations on the space-time differential forms like
$C^{0,2}$ and $\W_1^{0,3}$. At $M=4$ these equations are anticipated to be
off-shell constraints expressing some fields in terms
of derivatives of the others. In the language of unfolded
machinery this is equivalent to the statement that
the respective $\sigma_-$-cohomology groups analyzed in \cite{Gelfond:2013lba,gelap}
should be zero.

As explained in Section \ref{struc},
the appearance of the forms $\Ll$ makes the connections valued in the
$Q$-closed central elements
trivial, \ie Stueckelberg. For instance, $\Ll^2$ and $\Ll^4$ make
dynamically trivial $\W^{1,3}(0,\theta_x;0;0|x)$. In particular  the spin-one connection
valued in the center of the Chan-Paton group $U(n)$ of the original HS
theory  can be gauge fixed to zero in presence of $\Ll^2$. This does not mean,
 however, that spin-one massless modes disappear. They are still described
by the zero-form  $C^0$ obeying (\ref{dc}) (see also Section
\ref{Black holes}).

Finally, let us stress that the invariant functionals $S$ proposed in this paper
respect  the gauge transformations both in $x$-space and in the
space of spinorial coordinates $Z^A$ resulting by virtue of  (\ref{dzf})
 from formulae (\ref{dw}) and (\ref{db}) with  $\W_0^{0,1}$ (\ref{vac1}).
From the perspective of full nonlinear HS equations it is as important to
control the $Z^A$-gauge symmetry as that in $x$-space.

\subsection{Boundary functionals, parity, and $3d$ conformal HS theory }
\label{Parity}

As explained in Introduction, the local and non-local parts of the boundary
functionals are associated with different combinations of the coefficients in
(\ref{freel}), (\ref{Rxxzz}). Although these coefficients
can only be determined by the direct computation which is the subject of
\cite{DMV},  important piece of information can be deduced from the parity
properties of the theory.

{}From (\ref{xz}) it is clear that the parity transformation  $\zz\to-\zz$,
$\xx\to \xx$ is generated by the automorphism of the algebra that exchanges
left and right sectors, including the respective Klein
operators, \ie
\be
\theta^\ga, z^\ga, y^\ga, k \quad {\stackrel{P}{\Longleftrightarrow}}
 \quad \bar \theta^\dga, \bar z^\dga, \bar y^\dga,
\bar k\,.
\ee
For general $\eta$ in (\ref{etaB}),  HS equations (\ref{leq}) are not $P$-invariant.
However for the $A$-model with $\eta=1$ and $B$-model with $\eta=i$ they are
 provided that
\be
P\big (B(\theta;Z;Y;\ck|x)\big ) =\eta^2 B(P(\theta);P(Z);P(Y);P(\ck)|P(x))\,,
\ee
which implies, in particular, that the spin-zero modes of $\B$ describe scalars in the
$A$ model and pseudoscalars in the $B$-model \cite{Sezgin:2003pt}.

Since $\zz^{-1}d\zz$ is even under  $\zz\to-\zz$,
the non-zero contribution to the parity invariant functional (\ref{S}) comes from
the part $S^{loc}$  in (\ref{locnon}) that only contains boundary derivatives
(recall that an even combination of
$\zz$-derivatives  can be expressed via boundary derivatives by virtue of
the  field equations). Hence, for the $A$ and $B$-models  $S$ (\ref{S}) is
some gauge invariant boundary functional.
Since the $g$-dependent term in (\ref{leq}) is  $P$-invariant,
 the original bulk density form is invariant  under reflection of
 all bulk coordinates. As a result, upon the $\zz$ integration taking
 away one power of $\zz$, the  boundary functional should be odd under reflection
 of the boundary coordinates hence being of Chern-Simons type. The resulting
 gauge invariant local boundary functionals are conjectured to represent actions
 of $3d$ conformal HS theory.
Interestingly, our construction predicts two different actions for
$3d$ conformal HS theories associated with the $A$ and $B$ models.
As for the bulk theory, they  differ by the parity properties of the
scalar boundary current dual to the bulk scalar
field.

Naively, this consideration suggests that the nonlocal part of the boundary
functional in the $A$ and $B$ models is zero. This is not quite the case as we
explain now. To this end, consider the HS theory with  general $\eta$.
 Since, being invariant under the exchange of left and right sectors,
  the $g$-dependent term in (\ref{leq}) is  $P$-invariant, the whole setting is
 invariant under the $P$-transformation supplemented with $\eta\to\bar\eta$.
For our consideration it is essential that $L$ is evaluated at $Y=Z=0$ and
that the $g$-dependent term contains an additional factor of $k*\bar k$. This implies that
the computation in the dotted and undotted sectors are parallel except that in the
$g$-dependent contribution to the density  $k$ is replaced by $\bar k$ and vice versa.
As a result, with first on-shell theorem (\ref{FOST}),
the analogue of (\ref{freel}) has the structure
\be
L\sim \go (\eta\bar C + \bar \eta C)\,.
\ee

Setting  schematically
\be
\label{Rxxzz'}
R_{\xx\xx}\sim \eta e_\xx e_\xx C+
\bar \eta e_\xx e_\xx \bar C\q
R_{\xx\zz}\sim i\eta e_\zz e_\xx  C-i
\bar \eta e_\zz e_\xx \bar C\,
\ee
yields
\be
C\sim \bar \eta( R_{\xx\xx} -i R_{\xx\zz})\q \bar C\sim \eta
(R_{\xx\xx} +i R_{\xx\zz})\,,
\ee
For $\eta=\exp{i\varphi}$ this yields at the linearized level
\be
L \sim \go ( cos (2\varphi) R_{\xx\xx} - sin  (2\varphi) R_{\xx\zz} )\,,
\ee
\ie $S^{loc}$ contains the factor of $cos (2\varphi)$ while $S^{nloc}$
contains the factor of $sin (2\varphi)$.

Naively, this implies that, in accordance with the parity
analysis, $S^{nloc}$ vanishes at $\phi=0,\,\f{\pi}{2}$, \ie for $A$ and $B$ models.
However, to define both local and non-local functionals for the $A$ and $B$ models
 it makes sense to extract the factors of
$\cos( 2\varphi)$ and $\sin( 2\varphi)$ setting
\be
\label{A}
S_{A}^{loc}= S(0)\q S_A^{nloc} = \half \f{\p S(\varphi)}{\p \varphi} \Big|_{\varphi=0}\,,
\ee
\be
\label{B}
S_{B}^{loc}= S(\f{\pi}{2})\q S_B^{nloc} = \half \f{\p S(\varphi)}{\p \varphi}
\Big|_{\varphi=\f{\pi}{2}}\,.
\ee

Beyond the parity invariant HS models it is impossible to separate
the local and nonlocal parts of the gauge invariant functional $S$ (\ref{S}) since
only the full functional $S$ is gauge invariant. Indeed, the variation of the nonlocal
part can contain local terms compensating the nonzero gauge variation of the local
part. Only for  the $P$-invariant $A$ and $B$ models
it is possible to define the gauge invariant local boundary functionals
$S_{A,B}^{loc}$ to be identified with the actions of the boundary conformal
HS theory. (Note that our conclusions fit the identification of the action of the boundary
conformal HS theory with the local part of the boundary functional suggested
in \cite{Metsaev:2009ym} (see also \cite{Liu:1998bu}).)
On the other hand, the nonlocal functionals $S_{A,B}^{nloc}$
(\ref{A}), (\ref{B}) are guaranteed to be gauge invariant only up to
local terms resulting from the derivative of the gauge transformation of
$S^{loc}(\varphi)$ over $\varphi$, \ie the HS gauge symmetry of $S_{A,B}^{nloc}$
(and hence correlators) is respected up to local boundary terms.

It should be stressed that, due to differentiation over $\varphi$,  local
and nonlocal boundary functionals (\ref{A}) and (\ref{B}) have opposite
parity properties on the boundary. This implies in particular that the nonlocal
functional is parity even for parity-preserving bulk models.

\subsection{Black holes}
\label{Black holes}

The two-form part $\Ll^2$ of $\Ll$ in (\ref{leq})
is anticipated to support the BH charges.
In presence of $\Ll^2$, the spin-one sector of linearized Eq.~(\ref{FOST}) is
\be\label{1unf}
\dr_x\go^1(0;0;0|x)=\f{i}{8}\Big (\eta \overline H^{\dga\dgb}
\f{\p^2}{\p \by^\dga \p \by^\dgb} C^0(Y;\K|x )k +\bar \eta
H^{\ga\gb}  \f{\p^2}{\p y^\ga \p y^\gb} C^0(Y;\K|x )\bar k
\Big )\Big |_{Y=\K=0}-i\Ll^2\,,
\ee
where, abusing notations, we set $k\big |_{k=0} = 0$, $k^2\big |_{k=0} = 1$.
This implies
\be\label{2unf}
L^2=\f{1}{8}\Big (\eta \overline H^{\dga\dgb}
\f{\p^2}{\p \by^\dga \p \by^\dgb} C^0(Y;\K|x )k +\bar \eta
H^{\ga\gb}  \f{\p^2}{\p y^\ga \p y^\gb} C^0(Y;\K|x )\bar k
\Big )\Big |_{Y=\K=0}\,.
\ee
Let us stress that this provides a simple example of the situation with nonzero
$L^2$ implying that the supertrace in (\ref{lstr}) must be ill-defined. Other way around,
an assumption that the supertrace can be consistently regularized in (\ref{lstr})
 would imply that the
{\it r.h.s.} of Eq.~(\ref{2unf}) must be zero which is inconsistent since this is nothing else
as the {\it r.h.s.} of the equation expressing the zero-forms $C$ via the spin-one potential
$\go(0,0|x)$. In other words, if the supertrace in (\ref{lstr}) were well defined then Maxwell equations
for spin-one gauge potentials would not follow from usual HS equations (\ref{SS}).

As shown in  \cite{Didenko:2009tc}, a $4d$  GR BH solution is fully characterized by
a spin-one Papapetrou field \cite{Papa}. In terms of components of
the  field $C(Y|x)$ which extend the spin-two BH solution to all other
fields, the two-form field strength of the Papapetrou field  $\F$ is
\be\label{papa}
H^{\ga\gb}  C_{\ga\gb} +
\overline H^{\dga\dgb}\overline  C_{\dga\dgb}
 = M \F\,,
\ee
where $M$ is the BH mass  and zero-forms $C_{\ga\gb}$ and
$\overline C_{\dga\dgb}$ are self dual and anti-self dual components of the spin-one
field strength. (Recall that in this paper we use notations with anti-Hermitian potential
$\go^1(0;0;0|x)= i A(x)$, where $A(x)$ is the usual  electro-magnetic potential.)
 The Hodge dual two-form $\widetilde\F$  is
\be\label{hpapa}
 i\Big (
H^{\ga\gb}  C_{\ga\gb}
- \overline H^{\dga\dgb} \overline C_{\dga\dgb}
\Big ) = M \,\widetilde \F\,.
\ee
The Papapetrou field obeys the sourceless Maxwell equations everywhere except for
the singularity, \ie
both $\F$ and $\widetilde \F$ are closed,
\be
\dr_x \F =0\q \dr_x \widetilde \F=0\q x\neq 0\,.
\ee

For $\eta=\exp{[{i\varphi}]}$, Eq.~(\ref{2unf}) implies that
\be
L^2 = \f{1}{2} M\big (\cos (\varphi)\,\F +\sin(\varphi)\, {}\widetilde \F ) \,.
\ee

For the sake of simplicity in the sequel we consider the  case of
the Schwarzschild BH in GR leaving details of the general case to \cite{Didenko:2015pjo}.
The  Papapetrou two-form of the Schwarzschild BH is
\be
\F=\f{4}{r^2} dt dr\,,
\ee
where
$t$ and $r$ are the time and radial coordinates. Correspondingly,
\be
{}\widetilde \F= 4 d\Omega\,,
\ee
where $d\Omega$ is the angular two-form.
The properties of the form $M \widetilde \F$ suggest that, at least at the linearized
level in the HS theory, it should coincide with the two-form that supports the BH charge.
Indeed, at the horizon it has the form
\be
\widetilde \F= (2M)^{-2} V_H\,,
\ee
where $V_H$ is the horizon volume form. This gives
\be
L^2=\f{1}{2}\left ( \f{\sin(\varphi)}{4M} V_H +M \cos (\varphi)\F\right )\,.
\ee

For the Schwarzschild BH the second term does not contribute to the BH charge
resulting from the integration over space infinity while the first  gives
\be
 \int_\Sigma L^2 = \f{\sin(\varphi)}{8M} A_H\,,
\ee
where $A_H$ is the horizon area.
For the $A$-model with $\varphi=0$ this is  zero.
Analogously to the consideration of the boundary functional
in Section \ref{Parity}, a proper definition is
\be
\label{Q0}
Q(0) = \int_\Sigma  \f{\p L^2(\varphi)}{\p \varphi}\Big |_{\varphi =0}\,.
\ee
However application of this formula to a BH solution in the nonlinear
HS theory is not straightforward since exact HS BH solutions at $\varphi\neq 0$
are not yet available. We refer to \cite{Didenko:2015pjo} for a more general definition of
charges via variation over the modules  associated with the
topological fields of HS theory.

There are several reasons why $L^2(0)$ does not contribute to the BH
charge of the Schwarzschild BH while $Q(0)$ (\ref{Q0}) does. The simplest one
follows from the parity  analysis analogous to that of Section \ref{Parity}.
Another reason is that the Papapetrou field $\F$ is equivalent to the electromagnetic
field of a point-wise source. This implies that  equation (\ref{1unf}) admits a
solution with $\Ll^2=0$ and some $\go^1(0;0;0|x)$ regular at infinity, which is
just the Coulomb field. As a result, $L^2$ is exact at infinity and hence
cannot give a nonzero charge. On the other hand, $\tilde \F$ describes a
monopole solution. In this case, due to the Dirac string, the corresponding potential
$\go^1(0;0;0|x)$ is singular at $\Ll^2=0$. Hence, $L^2$, which is regular,  is closed
but not exact.

The  fact that the HS  theory  possesses a nontrivial on-shell closed form $L^2$  may look
surprising since it does not rely on a Killing symmetry of a particular solution,
holding for any solution. Indeed, no on-shell closed local density $L^2$ can be expected to exist
 in a  nonlinear on-shell theory in four dimensions.
The point  is  that the invariant  densities $L$ in the HS theory are
nonlocal and can involve infinitely
many derivatives of the dynamical fields with the coefficients containing inverse powers of
the cosmological constant. (The property that the cosmological constant is non-zero is
important  and the flat limit of $L^2$ is not obvious.)
Hence, the integral $Q=\int_{\Sigma^2} L^2(\phi)$ over some
surface $\Sigma^2$ may depend on the values of fields away from $\Sigma^2$.
Nevertheless, $L^2(\phi(x))$ is well-defined as a closed space-time two-form and
hence $Q$ is independent of local variations of $\Sigma^2$.
On the other hand,  evaluated for asymptotically
free theory at infinity, where  $L^2$ becomes asymptotically local, $Q$ correctly
reproduces usual asymptotic charges \cite{Didenko:2015pjo}.

Contracting Eq.~(\ref{dc}) with the time-like Killing vector $\xi^\un$ and using that
\be
\xi^\un \f{\p}{\p x^\un } \Big |_H = \f{\p}{\p t}
\Big |_H\q \xi^\un e_\un^{\ga\dga} \Big |_H=0
\ee
we observe that the generalized HS Weyl tensors
in  the unfolded equations for fluctuations of massless fields at the horizon
of the Schwarzschild BH are $t$-independent. Hence,
from the point of view of the observer at infinity, $Q$ evaluated at $H$ is associated
with the lower-dimensional system of $t$-independent fluctuations.
The form of this system can, in principle, be derived via reduction of system (\ref{hss}),
(\ref{leq}). It is tempting to speculate that this scheme can lead to
the identification of a microscopic pattern of the problem in terms of $L^2$.

\subsection{Vacuum partition}
Property (\ref{cow}) has a consequence that the vacuum value
of the density form $L^4$ is zero
\be
\label{L0}
L^4_0 =0\,
\ee
implying that the vacuum partition function is trivial,
$
Z_0=\exp  - S_0 =1\,.
$
Naively, this is true  for any boundary geometry consistent with the vacuum connection
obeying (\ref{W0eq}), (\ref{W0}), including
$AdS_3$  or $S^3$ in the Euclidean case. This conclusion is apparently
in contradiction with the holographic expectation of matching
the boundary vacuum partition (see e.g. \cite{Klebanov:2011gs,Giombi:2014yra}
and references therein).

Here however is a subtlety.
Indeed, if the cohomology $H^4$ of the boundary extended by
 the  complexified
Poincar\'e coordinate is nonzero one can look for another vacuum solution
with nonzero $L_0^4\in H^4$ and appropriately adjusted vacuum two-form
$\B_0^2$ in (\ref{leq}). This is analogous to the BH analysis in the previous
section where the closed form $L^2$ was supported by the Hodge dual
of the Papapetrou field via (\ref{hpapa}) with  $C_{\ga\gb}$ and
$\overline C_{\dga\dgb}$ being components of the zero-form  $\B^0$.
Such $L_0^4$ will contribute to the vacuum partition function.
 Remaining arbitrary, its magnitude will affect the perturbative
analysis becoming an essential parameter of the model analogous to the BH
mass. Careful analysis of this issue demands in particular an appropriate
reformulation of the  Poincar\'e-type foliation of the bulk space.
This is another interesting direction for the future work, being beyond the
scope of this paper. As an example, we consider below a particular realization of the
topological mechanism originating from the standard low-order frame-like
HS action \cite{Vasiliev:1986td}.

A typical
HS action \cite{Vasiliev:1986td} allowing a nonlinear deformation \cite{Fradkin:1987ks}
differs from the standard Fronsdal action \cite{Frhs,Frfhs} by a topological term.
In the spin-two gravitational sector this is  the MacDowell-Mansouri action
\cite{MacDowell:1977jt}
\begin{equation}
\label{MM}
S^{MM}=\frac{i}{4\kappa^2 \lambda^2}\int_{M^4} ( R_{\ga\gb}R^{\ga\gb}-
\bar R_{\dga\dgb}\bar R^{\dga\dgb})\,,
\end{equation}
where
\be
\label{nR1}
R_{\ga \gb}=\R_{\ga \gb} +\lambda^2\, e_{\ga}{}^{\dot{\delta}}e_{\gb \dot{\delta}}\q
\R_{\ga \gb}:= \dr_x\omega_{\ga \gb} +\omega_{\ga}{}^\gamma
 \omega_{\gb \gamma}
\,,
\ee
\be
\label{'}
\bar{R}_{{\dga} {\dgb}}=\bar{\R}_{{\dga} {\dgb}}
+\lambda^2\,
e^\gamma{}_{\dot{\alpha}}  e_{\gamma \dot{\gb}}\q
\bar{\R}_{{\dga} {\dgb}}:=\dr_x\bar{\omega}_{{\dga}
\dot{\gb}} +\bar{\omega}_{\dot{\alpha}}{}^{\dot{\gamma}}
\bar{\omega}_{\dot{\gb} \dot{\gamma}}\,
\ee
are the Lorentz components of the Riemann tensor shifted by the cosmological term,
which are defined in terms of vierbein $e^{\ga\dga}$ and Lorentz connection
$\go^{\ga\gb}$, $\bar \go^{\dga\dgb}$.
Along with the torsion two-form
\begin{equation}
\label{nrt}
R_{\alpha \dot{\beta}} :=\dr_x e_{\alpha\dot{\beta}} +\omega_\alpha{}^\gamma
e_{\gamma\dot{\beta}} +\bar{\omega}_{\dot{\beta}}{}^{\dot{\delta}}
 e_{\alpha\dot{\delta}}\,
\end{equation}
 they are components of the $sp(4)$ curvature
\be
R(Y|x):= \dr_x W(Y|x) + W(Y|x)*W(Y|x)\,,
\ee
\be
R(Y|x) = \f{i}{2}\big (R_{\ga \gb} y^\ga y^\gb + \bar{R}_{{\dga} {\dgb}} \bar y{}^\dga \bar y{}^\dgb
+ 2 R_{\alpha \dot{\beta}} y^\ga \bar y{}^\dgb \big )\,.
\ee

A locally $AdS_4$ space obeys  flatness equation (\ref{W0eq})
$
R(W_0)=0\,.
$
Hence, in accordance  with (\ref{L0}),
the MacDowell-Mansouri action is zero on any locally $AdS$ bulk.

The MacDowell-Mansouri action differs from
the Einstein-Hilbert action by the Gauss-Bonnet topological term. Indeed,
using (\ref{nR1}), (\ref{'}) we observe  that
\begin{equation}
\label{MM1}
S^{MM}= S^{top} +S^{EH} +S^c\,,
\end{equation}
where
\be
\label{top}
S^{top} = \frac{i}{4\kappa^2 \lambda^2}\int_{M^4} ( \R_{\ga\gb}\R^{\ga\gb}-
\bar\R_{\dga\dgb}\bar\R^{\dga\dgb})\,,
\ee
\be
\label{EH}
S^{EH} = \frac{i}{2\kappa^2 }\int_{M^4} ( e_{\ga}{}^{\dot{\delta}}
e_{\gb \dot{\delta}}\R^{\ga\gb}-
e^\gamma{}_{\dot{\alpha}}  e_{\gamma \dot{\gb}} \bar\R^{\dga\dgb})\,,
\ee
\be
\label{c}
S^{c} = \frac{i\lambda^2}{4\kappa^2 }\int_{M^4} ( e_{\ga}{}^{\dot{\delta}}
e_{\gb \dot{\delta}}e^{\ga}{}^{\dot{\gamma}}
e^\gb{}_{ \dot{\gamma}}-
e^\gamma{}_{\dot{\alpha}}  e_{\gamma \dot{\gb}}
e^\delta{}^{\dot{\alpha}}  e_{\delta}{}^{\dot{\gb}})\,.
\ee
Upon imposing the zero-torsion condition $R_{\ga\dga}=0$, which is one
of the field equations of the MacDowell-Mansouri action,  the two-forms
$\R_{\ga\gb}$ and $\bar\R_{\dga\dgb}$ describe the Riemann tensor.
Hence, $S^{EH}$ and $S^c$ are the Einstein-Hilbert action and the
cosmological term, respectively. The action $S^{top}$ is topological describing the
Euler characteristic of $M^4$. Its
variation over $\go_{\ga\gb}$ and $\bar \go_{\dga\dgb}$ is zero.
Generally, the vacuum contribution of $S^{top}$  is
nonzero, precisely compensating that of the  Einstein-Hilbert action.

In  the $AdS_4$ HS theory,
the Gauss-Bonnet contribution extends to higher spins as follows
\be
\label{HSgb}
S^{top} = \frac{i a}{4\kappa^2 \lambda^2} \int_{M^4} str \Big ( R^l(y;\K|x)* R^l(y;\K|x) -
\bar R^r(\bar y;\K|x)*\bar  R^r(\bar y;\K|x)\Big )\,,
\ee
where
\be
R^l(y;\K|x) = \dr \go^1 (y,0;\K|x) + \go^1(y,0;\K|x)*\go^1 (y,0;\K|x)\,
\ee
is expressed in terms of the one-form   HS fields (\ref{go})
($\bar  R^r(\bar y;\K|x)$ is complex conjugate to $R^l(y;\K|x)$).

Since,  at least in the lowest order,
the gauge invariant HS action enjoys the MacDowell-Mansouri form this can explain the compensation of the vacuum contribution
to the partition function in the gauge-invariant HS theory.
More generally,  possible contribution of the topological terms makes the vacuum
contribution to the action undetermined.

The Gauss-Bonnet Lagrangian provides an example of a nonzero vacuum
density form $L_0^4$. For the vacuum solution obeying (\ref{W0eq}) it is
proportional to the volume form
\be
\label{vact}
L_0^4 = bi H_{\ga\gb}H^{\ga\gb}=-bi\overline H_{\dga\dgb}\overline H^{\dga\dgb}
\ee
with some coefficient $b$ and two-forms $H_{\ga\gb}$, $\overline H_{\dga\dgb}$
(\ref{HH}). The $\pi_I$ projection of the \rhs of the four-form analogue of the $L^4$-extension of
equation (\ref{FOST}) is
\be
L^4=\f{1}{8}\big (\eta \overline H^{\dga\dgb}
\f{\p^2}{\p \by^\dga \p \by^\dgb} C^2(0,\by;\K|x )k +\bar \eta
H^{\ga\gb}{}  \f{\p^2}{\p y^\ga \p y^\gb} C^2(y,0;\K|x )\bar k\big )\Big |_{Y=0}\,.
\ee
To compensate the term with $L_0^4$ (\ref{vact}) it suffices to set
\be
\label{c2}
C_0^2(Y) = b \,i (\bar \eta
\overline H^{\dga\dgb}
\by_\dga  \by_\dgb k - \eta H^{\ga\gb} y_\ga y_\gb \bar k)  \,
\ee
using that $\eta\bar \eta = 1$ and
\be
H^{\ga\gb} \overline H^{\dga\dgb}=0\,
\ee
as a consequence of the relations
\be
\label{eee}
e_{(\ga}{}^\dga e_\gb{}^\dgb e_{\gga)}{}^{\dot \gga}=0\q
e_{\ga}{}^{(\dga} e_\gb{}^\dgb e_{\gga}{}^{\dot \gga )}=0
\ee
expressing the fact that the symmetrization over, say, three undotted indices of the vierbeins
implies the antisymmetrization over the three dotted ones, which take only two values.

It is important that, by virtue of (\ref{eee}),
$C^2(Y)$ (\ref{c2}) is covariantly constant obeying
equation analogous to (\ref{dB1}) thus solving (\ref{hss}).
Moreover, it cannot be represented  in the
exact form, \ie as the covariant derivative of something else, thus being
cohomologically nontrivial.
In the conventional definition
of the boundary functional (\ref{Se}), for conformally flat $M^4$ with volume $V_{M^4}$
this  gives a contribution proportional to $a \f{\lambda^2}{\kappa^{2}} V_{M^4} $
which remains arbitrary.

For the prescription (\ref{S}), the integration is over $S^1\times \Sigma^3$
where $S^1$ is the cycle around infinity and $\Sigma^3$ is the boundary surface.
Though for $\Sigma^3=S^3$ the additional contribution is likely to be zero in the
gravity case since the Euler number of $S^1\times \S^3$ is zero it would be
interesting to see  directly for an appropriate Poincar\'e-type foliation whether or
not it affects topology of the extended boundary making it different from
$S^3\times S^1$. In any case, if the outlined cohomological mechanism gives a
non-zero contribution
to the vacuum partition, in the proposed approach
its value becomes a free parameter distinguishing between
different phases of the theory.

\section{$AdS_3$ HS theory}
\label{ads3}

The form of nonlinear field equations of the $AdS_3$ HS theory
\cite{Prokushkin:1998bq} is  analogous to (\ref{SS}), (\ref{SB}).
The field variables  $W(\theta_x,z;y;\psi_{1,2};k | x)$,
$B(z;y;\psi_{1,2};k | x)$ and $S_\ga(z;y;\psi_{1,2};k | x)$
 depend on the space-time coordinates $x^\un$ $(\un=0,1,2)$, auxiliary
commuting spinors $z_\ga$, $y_\ga$ $(\ga=1,2)$, a pair of Clifford
elements $\{\psi_i,\psi_j\}=2\delta_{ij}$ $(i=1,2)$ that commute with
all other generating elements, and  the Klein operator $k$
\be
\label{Klein} k^2=1\q
   ky_\ga=-y_\ga k\q kz_\ga=-z_\ga k\,.
\ee

In terms of the one-form connection
\be
\W= \dr_x+W +S\,,
\ee
the $3d$ nonlinear field equations take the form
\be
\label{SS3}
\W*\W= -i \delta^2(\theta) (1 +   B* k*\ups)
\,,
\ee
\be
\label{SB3}
\W*B=B*\W\,.
\ee

By analogy with the $AdS_4$ HS equations a natural goal  would be
to construct a three-form density. However, this is impossible because every
three-form in the two-dimensional twistor space is zero that leaves
no room for a term analogous to that with $\delta^4(\theta)$ in (\ref{leq}).
Without such a term it is not clear how to generate nontrivial higher differential forms
both in the $\theta_z$ and in the $\theta_x$ sector which eventually would give rise to
a nontrivial three-form density. Note that a constant term proportional to
$\delta^2(\theta) k*\ups$ is contained in (\ref{SS3}) as a constant part of $B$. The
respective coupling constant was shown in \cite{Prokushkin:1998bq} to be related to the
parameter of mass of the matter fields in the $3d$ HS theory.

The absence of a  three-form density in the
$3d$ HS theory may be related to the peculiarity of
two-dimensional boundary conformal theory exhibiting
the holomorphic factorization. We conjecture that the
appropriate invariant functional in the $3d$ HS theory is supported by
 a two-form $L^2(\theta_x |x)$ resulting from the following generalization of (\ref{SS3})
\be
\label{SS3L}
\W*\W= -i \left (\delta^2(\theta) (1 +   B* k*\ups) +\Ll^2(\theta_x|x)\, I\right )\q \dr_x \Ll^2(\theta_x|x)=0\,.
\ee
The part of $L^2$, that  contributes to the generating functional (\ref{S}),
is
\be
L = d\zz (d\xx L_{\zz \xx} + d{\bar \xx} \bar L_{\zz \bar \xx})\,,
\ee
where $\xx $ and $\bar \xx$ are complex coordinates of the two-dimensional boundary.
The invariant functionals $S=\int L^2$ should result from the integration
over $S^1\times \Sigma$ where $S^1$ is a contour around the $AdS_3$ infinity  $\zz=0$
while $\Sigma $ is a complex  curve on the  $2d$ boundary.
Since  $L^2$ is closed, the result is independent of local variations
of $\Sigma$. Hence, for a Riemann surface $\Sigma$, so defined $S$ will only
depend on its genus.

This conjecture can be checked  using the analysis
of the boundary behavior in $AdS_3$ of \cite{Vasiliev:2012vf}.
The same two-form density $L^2$ integrated over a different cycle
surrounding  the BH singularity is anticipated to describe
charges of the BTZ-like BH solutions \cite{Didenko:2006zd,Ammon:2012wc} in $3d$
HS theories. We hope to consider these problems in more detail elsewhere.

Analogously to the $4d$ case, the extension of the set of fields by the two-form $\Ll^2$
makes the one-form $\go:=\W(0,\theta_x;0;0;0;0|x)$ dynamically trivial.
The difference is that in the $3d$ theory the gauge fields are of Chern-Simons
type admitting no non-zero  on-shell curvatures analogous to the Weyl-like tensors
in $4d$ HS theory.
This implies that there is no room for the zero-forms $C$ in the $3d$ First On-Shell
Theorem which has the form
\be
\dr_x\go(x) = \Ll^2+\ldots\,,
\ee
where ellipses denotes nonlinear corrections. As a result,
 $L^2$  starts from  quadratic terms in the zero-forms $C$ describing
 matter fields of spins 0 and 1/2. This is just appropriate for the
 generating functional of the boundary correlators.

\section{Conclusion}
\label{Conclusion}

The construction of invariant  functionals $S$ in HS  theory proposed in this paper
associates them with central  elements of the HS algebra. $S$
are integrals of  space-time differential forms $L$ that are closed by
virtue of the HS field equations.
The  densities $L$ are specific fields in the
extended unfolded system of HS equations, which are expressed  by this  system
in terms  of the other fields. Since the gauge transformation of $L$ has the  form
$\delta L = \dr \chi$ where $\chi$ is a function (\ref{gtl})
of other fields and gauge parameters in the system, the functionals $S$ are
gauge invariant. The new element
of our construction is that nontrivial functionals $S$ are
conjectured to be supported by such densities
$L$ that cannot be represented in the form of supertrace of some pre-density,
\ie  $L\neq str (L')$ for any $L'$ built from the HS gauge fields. In this respect
our proposal differs from most of other proposals in the literature where invariant
functionals are searched in the form of supertrace of a pre-density $L'$.

The closely related  property is that the invariant functionals proposed in this
paper are not local, containing infinite expansions in powers of derivatives.
Since the unbroken phase of the HS theories is anticipated to describe
physics at ultrahigh (transPlanckian) energies, such theories should be
non-local  one way or another. It is important  to specify the degree of nonlocality
in such theories. In \cite{Vasiliev:2015wma} a criterion was suggested distinguishing
between local,  minimally nonlocal and strongly nonlocal functionals. The degree of
nonlocality in the HS theory is minimally nonlocal.

The density forms $L_i$ are associated with certain central elements $c_0^i$
of the HS algebra. Introduction of the density forms $L_i$ has a consequence
  that the  fields of the original HS theory proportional to the central elements $c^i_0$
disappear becoming Stueckelberg with respect to the gauge symmetries
associated with the differential forms $\Ll_i$ underlying the construction of $L_i$.
For instance, the spin-one connection that carries no
color indices disappears from the   $3d$ HS theory due to the gauge symmetry
of the  two-form $\Ll^2$.

Our scheme is coordinate independent being applicable to configurations of any topology.
To be nontrivial, invariant actions have to be integrated over noncontractible cycles.
In the on-shell case, one option is to integrate a density $d$-form  $L^d$
over  $S^1\times \Sigma^{d-1}$ where $\Sigma^{d-1}$ is a
$(d-1)$-dimensional boundary of the $d$-dimensional bulk space while $S^1$ is a circle
around the infinite point $\zz=0$ on the complex plane of the complexified Poincar\'e
coordinate. The respective action (\ref{S}) is conjectured to give rise to the generating functional
of  correlators of the boundary theory. Standard functional (\ref{Se}) can also be
considered. Explicit check of whether  the proposed functional properly reproduces boundary
correlators  will be reported in \cite{DMV}. The analysis of
this paper shows however that some of the generating functionals for nonlocal
contributions to the boundary correlators in the HS theory
should be associated with derivatives  $\f{\p L^4(\varphi)}{\p \varphi}\Big|_{\varphi =0}$
rather than $L^4(0)$, where $\varphi$ is the phase parameter distinguishing between
different HS models. In these cases $L^4(0)$ describes the
Lagrangian of the boundary conformal HS theory
that only gives local contribution to the correlators. Note that
with this definition local boundary functionals are parity odd in agreement with
the expectation that $3d$ conformal HS theory should have Chern-Simons form, while
the nonlocal ones are parity even.

Hopefully, the construction of the boundary functional in the form (\ref{S}) may have
applicability beyond   HS theories. The peculiar property that
the integration is over the region beyond $AdS$ infinity may have something to do with
the classical to quantum transmutation in the $AdS/CFT$ holography being
somewhat reminiscent of  quantum tunneling allowing to reach configurations
unreachable in classical physics.

Another problem is to evaluate invariants associated with $(d-2)$--forms as integrals
over lower-dimensional surfaces surrounding a BH singularity. Invariants  of
this class, including derivatives $\f{\p L^2(\varphi)}{\p \varphi}\Big|_{\varphi =0}$,
 are conjectured to describe the BH charges in HS theory.
It is tempting to speculate that the proposed approach may provide tools for a microscopic interpretation of the
BH entropy in terms of the unfolded system associated with the pushforward of the original
system to the horizon. An intriguing
point is that the BH problem turns out to be analogous to the $AdS/CFT$ problem
since the BH solutions in the HS theory \cite{Didenko:2009td,Iazeolla:2011cb}
are based on the Fock vacua in the twistor space analogous to the Fock vacua
(\ref{fock}), (\ref{bfock}) which determine the boundary behavior of the bulk fields
\cite{Vasiliev:2012vf}. In fact, the analysis of BH physics is in a certain sense
technically simpler than of the boundary correlators since in the former case
nontrivial contributions start from the first order while in the latter
 from the second. Specifically, as shown in Section \ref{Black holes},
in the $4d$ HS theory $L^2$ identifies with the spin-one field strength of the
Papapetrou field \cite{Papa}.

It should be stressed that the  existence of the form $L^2$ closed on the HS
field equations is possible because, away from the free field limit,
$L^2$ is a nonlocal functional of the dynamical fields. Such objects
naturally appear in the HS theory formulated in the $AdS$ space but
can hardly be introduced in  conventional local theories in  Minkowski space.

In this paper we consider the on-shell HS systems in $AdS_4$ and $AdS_3$,
formulated in terms of spinorial star-product algebras.
An interesting peculiarity of the on-shell spinorial HS theory in $AdS_3$
is that the density form
of  maximal degree in this theory is a two-form. From the $AdS_3/CFT_2$
correspondence perspective
this implies that it should be integrated over a one-dimensional surface
of the boundary times the circle around infinity. This picture matches
 holomorphicity of two-dimensional conformal theories. One of the most interesting
 problems for the future is  to see details of this mechanism in the $AdS_3/CFT_2$
HS holography.

The proposed construction raises many  questions for the further work.
Our approach applies to both  on-shell and  off-shell
unfolded systems. An interesting problem is
to construct on-shell and off-shell invariants of the vectorial HS theories
of \cite{Vasiliev:2003ev}.
This requires analysis of $Q$-cohomology in these theories  as well
as the proper extension of the construction of functional classes of
\cite{Vasiliev:2015wma} which is more subtle because the
HS algebra underlying vectorial HS theory is not freely generated, resulting
from quotiening  certain constraints.
Another interesting problem is to work out the form of the on-shell densities in the
conventional lower-spin theories. Also it is  important to investigate
more carefully the structure of boundary singularities associated with the Fock behavior
at $\zz\to 0$ initiated in Section \ref{Fock}.

\section*{Acknowledgments}
I am grateful to Abbay Ashtekar, Glenn Barnich, Andrey Barvinsky,
 Olga Gelfond, Maxim Grigoriev,  Murat Gunaydin,
Olaf Hohm, Carlo Iazeolla, Hong Liu, Ruslan Metsaev, Benght Nilsson,
Eric Perlmutter,  Anastasios Petkou, Per Sundell and Arkady Tseytlin  for stimulating
discussions  and especially to Igor Klebanov also for triggering this research.
I am particularly grateful to Slava Didenko and Nikita Misuna  for many most useful comments and discussions.
This research was supported by the Russian Science Foundation Grant No 14-42-00047.


\begin{thebibliography}{99}
\parindent=0pt
\parskip=0pt


\bibitem{Maldacena:1997re}
J.~M.~Maldacena,
  Adv.\ Theor.\ Math.\ Phys.\  {\bf 2} (1998) 231
  [Int.\ J.\ Theor.\ Phys.\  {\bf 38} (1999) 1113]
  [arXiv:hep-th/9711200].
\bibitem{Gubser:1998bc}
S.~S.~Gubser, I.~R.~Klebanov and A.~M.~Polyakov,
  Phys.\ Lett.\  B {\bf 428}, 105 (1998)
  [arXiv:hep-th/9802109].

\bibitem{Witten:1998qj}
E.~Witten,
  Adv.\ Theor.\ Math.\ Phys.\  {\bf 2}, 253 (1998)
  [arXiv:hep-th/9802150].

\bibitem{D'Hoker:1998tz}
  E.~D'Hoker, D.~Z.~Freedman and W.~Skiba,
  Phys.\ Rev.\ D {\bf 59} (1999) 045008
  [hep-th/9807098].

\bibitem{Frhs}
C.~Fronsdal, {\it Phys. Rev.\/} D {\bf 18} (1978) 3624; D {\bf 20}
              (1979) 848.
\bibitem{Frfhs}
              J.~Fang and C.~Fronsdal, {\it Phys. Rev.\/} D {\bf
              18} (1978) 3630; D {\bf 22} (1980) 1361.

\bibitem{Bengtsson:1983pd}
A.~K.~H. Bengtsson, I.~Bengtsson, and L.~Brink, ``{Cubic interaction terms for
  arbitrary spin},''
{{\em Nucl. Phys.}
  {\bfseries B227} (1983) 31}.

\bibitem{Berends:1984wp}
F.~A. Berends, G.~J.~H. Burgers, and H.~Van~Dam, ``{On spin three
  selfinteractions},''
{{\em Z. Phys.} {\bfseries C24}
  (1984) 247--254}.


\bibitem{Fradkin:1987ks}
E.~S. Fradkin and M.~A. Vasiliev, ``{On the Gravitational Interaction of
  Massless Higher Spin Fields},''
{{\em Phys. Lett.}
  {\bfseries B189} (1987) 89--95}.

\bibitem{Vasilev:2011xf}
  M.~A.~Vasiliev,
  Nucl.\ Phys.\ B {\bf 862} (2012) 341
  [arXiv:1108.5921 [hep-th]].

\bibitem{Joung:2011ww}
  E.~Joung and M.~Taronna,
  Nucl.\ Phys.\ B {\bf 861} (2012) 145
  [arXiv:1110.5918 [hep-th]].

\bibitem{Bekaert:2014cea}
  X.~Bekaert, J.~Erdmenger, D.~Ponomarev and C.~Sleight,
  JHEP {\bf 1503} (2015) 170
  [arXiv:1412.0016 [hep-th]].



\bibitem{Boulanger:2011dd}
  N.~Boulanger and P.~Sundell,
  J.\ Phys.\ A  {\bf 44} (2011) 495402
  [arXiv:1102.2219 [hep-th]].

\bibitem{Sezgin:2011hq}
  E.~Sezgin and P.~Sundell,
  JHEP {\bf 1207} (2012) 121
  [arXiv:1103.2360 [hep-th]].

\bibitem{Klebanov:2002ja}
  I.~R.~Klebanov and A.~M.~Polyakov,
  Phys.\ Lett.\  B {\bf 550} (2002) 213
  [arXiv:hep-th/0210114].

\bibitem{Leigh:2003gk}
  R.~G.~Leigh and A.~C.~Petkou,
  JHEP {\bf 0306} (2003) 011
  [hep-th/0304217].


\bibitem{Sezgin:2003pt}
  E.~Sezgin and P.~Sundell,
  JHEP {\bf 0507} (2005) 044
  [arXiv:hep-th/0305040].

\bibitem{Giombi:2009wh}
  S.~Giombi and X.~Yin,
  JHEP {\bf 1009} (2010) 115
  [arXiv:0912.3462 [hep-th]].




\bibitem{Vasiliev:2012vf}
  M.~A.~Vasiliev,
  J.\ Phys.\ A {\bf 46} (2013) 214013
  [arXiv:1203.5554 [hep-th]].


\bibitem{Maldacena:2012sf}
  J.~Maldacena and A.~Zhiboedov,
  Class.\ Quant.\ Grav.\  {\bf 30} (2013) 104003
  [arXiv:1204.3882 [hep-th]].

\bibitem{Giombi:2012ms}
  S.~Giombi and X.~Yin,
  J.\ Phys.\ A {\bf 46} (2013) 214003
  [arXiv:1208.4036 [hep-th]].

\bibitem{Colombo:2012jx}
  N.~Colombo and P.~Sundell,
  arXiv:1208.3880 [hep-th].

\bibitem{Didenko:2012tv}
  V.~E.~Didenko and E.~D.~Skvortsov,
  JHEP {\bf 1304} (2013) 158
  [arXiv:1210.7963 [hep-th]].



\bibitem{Jevicki:2012fh}
  A.~Jevicki, K.~Jin and Q.~Ye,
  J.\ Phys.\ A {\bf 46} (2013) 214005
  [arXiv:1212.5215 [hep-th]].

\bibitem{Giombi:2013fka}
  S.~Giombi and I.~R.~Klebanov,
  JHEP {\bf 1312} (2013) 068
  [arXiv:1308.2337 [hep-th]].

\bibitem{Giombi:2014yra}
  S.~Giombi, I.~R.~Klebanov and A.~A.~Tseytlin,
  Phys.\ Rev.\ D {\bf 90} (2014) 024048
  [arXiv:1402.5396 [hep-th]].

\bibitem{Beccaria:2014jxa}
  M.~Beccaria, X.~Bekaert and A.~A.~Tseytlin,
  JHEP {\bf 1408} (2014) 113
  [arXiv:1406.3542 [hep-th]].

\bibitem{Koch:2014aqa}
  R.~d.~M.~Koch, A.~Jevicki, J.~P.~Rodrigues and J.~Yoon,
  J.\ Phys.\ A {\bf 48} (2015) no.10,  105403
  [arXiv:1408.4800 [hep-th]].


\bibitem{Giombi:2014xxa}
  S.~Giombi and I.~R.~Klebanov,
  JHEP {\bf 1503} (2015) 117
  [arXiv:1409.1937 [hep-th]].

\bibitem{Beccaria:2014xda}
  M.~Beccaria and A.~A.~Tseytlin,
  JHEP {\bf 1411} (2014) 114
  [arXiv:1410.3273 [hep-th]].


\bibitem{Barvinsky:2014kta}
  A.~O.~Barvinsky,
  J.\ Exp.\ Theor.\ Phys.\  {\bf 120} (2015) 3,  449
  [arXiv:1410.6316 [hep-th]].

\bibitem{Henneaux:2010xg}
  M.~Henneaux and S.~J.~Rey,
  JHEP {\bf 1012} (2010) 007
  [arXiv:1008.4579 [hep-th]].

\bibitem{Campoleoni:2010zq}
  A.~Campoleoni, S.~Fredenhagen, S.~Pfenninger and S.~Theisen,
  JHEP {\bf 1011} (2010) 007
  [arXiv:1008.4744 [hep-th]].

\bibitem{Gaberdiel:2010pz}
  M.~R.~Gaberdiel and R.~Gopakumar,
  Phys.\ Rev.\  D {\bf 83} (2011) 066007
  [arXiv:1011.2986 [hep-th]].

\bibitem{more} M.~A.~Vasiliev, {\it Phys. Lett.}  B {\bf 285} (1992) 225.

\bibitem{Arutyunov:1998ve}
 G.~E.~Arutyunov and S.~A.~Frolov,
 Nucl.\ Phys.\  B {\bf 544}, 576 (1999)
 [arXiv:hep-th/9806216].


\bibitem{Metsaev:2009ym}
 R.~R.~Metsaev,
 Phys.\ Rev.\ D {\bf 81}, 106002 (2010)
 [arXiv:0907.4678 [hep-th]].

\bibitem{Metsaev:2014vda}
  R.~R.~Metsaev,
  Theor.\ Math.\ Phys.\  {\bf 181} (2014) no.3,  1548
  [arXiv:1407.2601 [hep-th]].


\bibitem{Castellani:1981um}
  L.~Castellani, P.~Fre and P.~van Nieuwenhuizen,
  Annals Phys.\  {\bf 136} (1981) 398.

\bibitem{Castellani:1991ev}
  L.~Castellani, R.~D'Auria and P.~Fre,
  Singapore, Singapore: World Scientific (1991) 1375-2162

\bibitem{Hu:2015cwa}
  S.~Hu and T.~Li,
  JHEP {\bf 1510} (2015) 019
  [arXiv:1501.02322 [hep-th]].


\bibitem{Pope:1989vj}
  C.~N.~Pope and P.~K.~Townsend,
  Phys.\ Lett.\ B {\bf 225} (1989) 245.


\bibitem{Fradkin:1989xt}
  E.~S.~Fradkin and V.~Y.~Linetsky,
  Mod.\ Phys.\ Lett.\  A {\bf 4} (1989) 731
  [Annals Phys.\  {\bf 198} (1990) 293].

\bibitem{Horne:1988jf}
  J.~H.~Horne and E.~Witten,
  Phys.\ Rev.\ Lett.\  {\bf 62} (1989) 501.


\bibitem{Nilsson:2013tva}
  B.~E.~W.~Nilsson,
  JHEP {\bf 1509} (2015) 078
  [arXiv:1312.5883 [hep-th]].


\bibitem{Witten:2003ya}
  E.~Witten,
  In *Shifman, M. (ed.) et al.: From fields to strings, vol. 2* 1173-1200
  [hep-th/0307041].

\bibitem{Leigh:2003ez}
  R.~G.~Leigh and A.~C.~Petkou,
  JHEP {\bf 0312} (2003) 020
  [hep-th/0309177].

\bibitem{Didenko:2015cwv}
  V.~E.~Didenko, N.~G.~Misuna and M.~A.~Vasiliev,
  JHEP {\bf 1607} (2016) 146
  [arXiv:1512.04405 [hep-th]].

\bibitem{DMV}
V.~E.~Didenko, N.~G.~Misuna and M.~A.~Vasiliev,  work in progress.

\bibitem{Wald:1993nt}
  R.~M.~Wald,
  Phys.\ Rev.\ D {\bf 48} (1993) 3427
  [gr-qc/9307038].

\bibitem{Didenko:2015pjo}
  V.~E.~Didenko, N.~G.~Misuna and M.~A.~Vasiliev,
  arXiv:1512.07626 [hep-th].

\bibitem{Vasiliev:2015wma}
  M.~A.~Vasiliev,
  JHEP {\bf 1506} (2015) 031
  [arXiv:1502.02271 [hep-th]].

\bibitem{Barnich:2001jy}
  G.~Barnich and F.~Brandt,
  Nucl.\ Phys.\ B {\bf 633} (2002) 3
  [hep-th/0111246].

\bibitem{Barnich:2005bn}
  G.~Barnich, N.~Bouatta and M.~Grigoriev,
  JHEP {\bf 0510} (2005) 010
  [hep-th/0507138].

\bibitem{Didenko:2009td}
  V.~E.~Didenko and M.~A.~Vasiliev,
  Phys.\ Lett.\ B {\bf 682} (2009) 305
   [Erratum-ibid.\ B {\bf 722} (2013) 389]
  [arXiv:0906.3898 [hep-th]].


\bibitem{Iazeolla:2011cb}
  C.~Iazeolla and P.~Sundell,
  JHEP {\bf 1112} (2011) 084
  [arXiv:1107.1217 [hep-th]].

\bibitem{Bourdier:2014lya}
  J.~Bourdier and N.~Drukker,
  JHEP {\bf 1504} (2015) 097
  [arXiv:1411.7037 [hep-th]].


\bibitem{Strominger:1996sh}
  A.~Strominger and C.~Vafa,
  Phys.\ Lett.\ B {\bf 379} (1996) 99
  [hep-th/9601029].




\bibitem{BTZ}
M. Banados, C. Teitelboim and J. Zanelli, \textit{Phys.Rev.Lett.}
{\bf 69} (1992) 1849, hep-th/9204099

\bibitem{Didenko:2006zd}
  V.~E.~Didenko, A.~S.~Matveev and M.~A.~Vasiliev,
  Theor.\ Math.\ Phys.\  {\bf 153} (2007) 1487
   [Teor.\ Mat.\ Fiz.\  {\bf 153} (2007) 158]
  [hep-th/0612161].

\bibitem{Ammon:2012wc}
  M.~Ammon, M.~Gutperle, P.~Kraus and E.~Perlmutter,
  J.\ Phys.\ A {\bf 46} (2013) 214001
  [arXiv:1208.5182 [hep-th]].

\bibitem{Ann}
M.~A.~Vasiliev,
Ann. Phys. (NY) {\bf 190} {(1989)} {59}.


\bibitem{Sullivan}
D. Sullivan, Publ.\ Math.\ IH\'ES {\bf 47} (1977) 269.


\bibitem{FDA}
R.D'Auria and P.~Fre,
 Nucl. Phys.  B {\bf 201} (1982) 101 [Erratum-ibid. B {\bf 206}
(1982) 496.]

\bibitem{FDA1}
P. van Nieuwenhuizen, ``Free Graded Differential Superalgebras,'' in M. Serdaroglu and E. In\"{o}n\"{u} ed., {\it Group Theoretical Methods
in Physics: Proceedings}, Lecture Notes in Physics, Vol.180
(Springer-Verlag, 1983).

\bibitem{FDA2}
R.~D'Auria, P.~Fre, P.~K.~Townsend and P.~van Nieuwenhuizen,
Ann. of Phys.\  {\bf 155} (1984) 423.

\bibitem{Bekaert:2005vh}
  X.~Bekaert, S.~Cnockaert, C.~Iazeolla and M.~A.~Vasiliev,
  arXiv:hep-th/0503128.

\bibitem{act} M.~A.~Vasiliev,
{\it  Int.J.Geom.Meth.Mod.Phys.} {\bf 3} (2006) 37 [hep-th/0504090].

\bibitem{Douglas:2010rc}
  M.~R.~Douglas, L.~Mazzucato and S.~S.~Razamat,
  Phys.\ Rev.\ D {\bf 83} (2011) 071701
  [arXiv:1011.4926 [hep-th]].

\bibitem{Sachs:2013pca}
  I.~Sachs,
  Phys.\ Rev.\ D {\bf 90} (2014) 8,  085003
  [arXiv:1306.6654 [hep-th]].

\bibitem{Leigh:2014qca}
  R.~G.~Leigh, O.~Parrikar and A.~B.~Weiss,
  Phys.\ Rev.\ D {\bf 91} (2015) no.2,  026002
  [arXiv:1407.4574 [hep-th]].

\bibitem{Mintun:2014gua}
  E.~Mintun and J.~Polchinski,
  arXiv:1411.3151 [hep-th].


\bibitem{de Boer:1999xf}
  J.~de Boer, E.~P.~Verlinde and H.~L.~Verlinde,
  JHEP {\bf 0008} (2000) 003
  [hep-th/9912012].


\bibitem{Sezgin:2005pv}
  E.~Sezgin and P.~Sundell,
  Nucl.\ Phys.\ B {\bf 762} (2007) 1
  [hep-th/0508158].

\bibitem{Gelfond:2013xt}
  O.~A.~Gelfond and M.~A.~Vasiliev,
  Nucl.\ Phys.\ B {\bf 876} (2013) 871
  doi:10.1016/j.nuclphysb.2013.09.001
  [arXiv:1301.3123 [hep-th]].


\bibitem{V3}     M.~A.~Vasiliev, {\it Fortschr. Phys.\/} {\bf 36} (1988) 33.

\bibitem{KV}  S.~E.~Konstein and M.~A.~Vasiliev, {\it Nucl. Phys.\/} {\bf B331} (1990) 475.

\bibitem{Vasiliev:1999ba}
  M.~A.~Vasiliev,
  arXiv:hep-th/9910096.

\bibitem{Chang:2012kt}
  C.~M.~Chang, S.~Minwalla, T.~Sharma and X.~Yin,
  J.\ Phys.\ A {\bf 46} (2013) 214009
  [arXiv:1207.4485 [hep-th]].

\bibitem{Vasiliev:2003ev}
  M.~A.~Vasiliev,
  Phys.\ Lett.\ B {\bf 567} (2003) 139
  [hep-th/0304049].


\bibitem{333}
M.~A.~Vasiliev, { Nucl.Phys.} B {\bf 793} (2008) 469, {\tt
arXiv:0707.1085 [hep-th]}.

\bibitem{Gelfond:2013lba}
  O.~A.~Gelfond and M.~A.~Vasiliev,
  JHEP {\bf 1610} (2016) 067
  [arXiv:1312.6673 [hep-th]].

\bibitem{gelap}
  O.~A.~Gelfond and M.~A.~Vasiliev, in preparation.



\bibitem{Liu:1998bu}
 H.~Liu and A.~A.~Tseytlin,
 Nucl.\ Phys.\ B {\bf 533}, 88 (1998)
 [arXiv:hep-th/9804083].


\bibitem{Didenko:2009tc}
  V.~E.~Didenko, A.~S.~Matveev and M.~A.~Vasiliev,
  arXiv:0901.2172 [hep-th].


\bibitem{Papa}
A. Papapetrou, \emph{Stationary gravitational fields with axial symmetry},
Ann. Inst. H. Poincar\'e {\bf A4} (1966) 83

\bibitem{Klebanov:2011gs}
  I.~R.~Klebanov, S.~S.~Pufu and B.~R.~Safdi,
  JHEP {\bf 1110} (2011) 038
  [arXiv:1105.4598 [hep-th]].

\bibitem{Vasiliev:1986td}
  M.~A.~Vasiliev,
  Fortsch.\ Phys.\  {\bf 35} (1987) 741
   [Yad.\ Fiz.\  {\bf 45} (1987) 1784].


\bibitem{MacDowell:1977jt}
S.~W. MacDowell and F.~Mansouri,
{ Phys. Rev. Lett.
  {\bfseries 38} (1977) 739}.


\bibitem{Prokushkin:1998bq}
  S.~F.~Prokushkin and M.~A.~Vasiliev,
  Nucl.\ Phys.\ B {\bf 545} (1999) 385
  [hep-th/9806236].









\end{thebibliography}
\end{document}